\RequirePackage{fix-cm}

\documentclass[smallextended]{svjour3VISbi}       
\smartqed  
\usepackage{graphicx}
\usepackage{amssymb}

\newcommand{\Rc}{\mathcal{R}}
\newcommand{\Sc}{\mathcal{S}}

\newcommand{\bJ}{\mathbb{J}}

\newcommand{\Times}{\!\times\!}
\newcommand{\Cdot}{\!\cdot\!}
\newcommand{\Schrodinger}{Schr\"{o}dinger~}

 \journalname{Foundations of Physics (2019)  1-38,  in Press}
 \doi{DOI: 10.1007/s10701-019-00236-4}
\begin{document}

\title{Quaternion algebra on 4D superfluid quantum space-time. Gravitomagnetism}



\author{Valeriy I. Sbitnev
}


\institute{V. I. Sbitnev \at
              St. Petersburg B. P. Konstantinov Nuclear Physics Institute, NRC Kurchatov Institute, Gatchina, Leningrad district, 188350, Russia; \\
               Department of Electrical Engineering and Computer Sciences, University of California, Berkeley, Berkeley, CA 94720, USA\\ \\
              Tel.: +781-37-137944\\
              \email{valery.sbitnev@gmail.com}      
}

\date{\today}

\maketitle

\begin{abstract}
Gravitomagnetic equations result from applying quaternionic differential operators to the energy-momentum tensor. These equations are similar to the Maxwell's EM equations. Both sets of the equations are isomorphic after changing orientation of either the gravitomagnetic orbital force or the magnetic induction. The gravitomagnetic equations turn out to be parent equations generating the following set of equations: (a) the vorticity equation giving solutions of vortices with nonzero vortex cores and with infinite lifetime; (b) the Hamilton-Jacobi equation loaded by the quantum potential. This equation in pair with the continuity equation leads to getting the \Schrodinger equation describing a state of the superfluid quantum medium (a modern version of the old ether); (c) gravitomagnetic wave equations loaded by forces acting on the outer space. These waves obey to the Planck's law of radiation.


\end{abstract}

\section{\label{sec1}Introduction}

Roger Penrose in his earlier works~\cite{Penrose1967,Penrose1968} has proposed the formalism of twistor algebra where four parameters of the space-time, $t, x, y, z$, by amazing manner find accordance with spinor group SU(2). Twistor theory was originally proposed as a new geometric framework for physics that aims to unify general relativity and quantum mechanics~\cite{AtiyahEtAl2017}.
An unexpected interconnection between the geometry of Minkowski space-time and the twisting of space-time leads to a new view of the relation between quantum theory and space-time curvature. 
A basic type of the twistor with four complex components is given by a pair of two-component spinors, one of which defines the direction and the other defines its angular moment about the origin. Components of  Minkowski space are subjected to the complexification in order to use a powerful apparatus of the unitary unimodular group SU(2).

 Here we pass from the group SU(2) to its quaternion representation through the real $4\Times4$ matrices and transform Penrose's twistor ideas for using the Clifford algebra. In fact, we consider motions on the ten-dimensional space-time manifold by applying the 
Clifford algebra~\cite{Baylis1996,Girard2007,Lounesto2001,Hestenes2015}.
 It turns out, that the quaternion $4\times4$ matrices allow also to describe, in addition to the translation of the space-time, the appearance of torsion fields~\cite{Penrose1983,Rapoport2005a,Rapoport2005,Rapoport2007,Shipov1998}.
 Similar type of the torsion, the Cartan's torsion, which was put into oblivion, was  under consideration of \Schrodinger, Einstein, Dirac at developing a unified field theory~\cite{Goenner2004,Goenner2014,HehlObukhov2007}.
 
Clifford algebras are important mathematical tools in the knowledge of the subtle secrets of Nature. 
 This compliance is marked by many authors in special literature devoted to this
 issue~\cite{Baylis1996,Girard1984,Girard2007,Hestenes2015,Lounesto2001,Rapoport2005a,ScolariciSolombrino1995}. 
The quaternions serve as a tool for describing both translations in 4D space and spin rotations on 3D spheres of unit radius~\cite{AgamalyanEtAl1988}.
The Lorentz transformations of 4D space-time is described through the quaternion matrices as well~\cite{Sbitnev2018c}.
 Roger Penrose and Wolfgang Rindler devoted two books  to this issue, both entitled
 "Spinors and Space-Time"~\cite{PenroseRindler1984,PenroseRindler1986} in which the above were considered in details.

  Here we deal with the space densely filled by an incompressible quantum superfluid described as a 
  Bose-Einstein condensate~\cite{Huang2013,HuangEtAl2014,Volovik2013,SarkarEtAl2018}. At low temperature of the cosmic space, the vacuum energy density scales as non-relativistic matter~\cite{AlbaretiEtAl2014a,AlbaretiEtAl2014b}. In this perspective, computations lead to the gravitomagnetic equations~\cite{MashhoonEtAl2001} strongly similar to the Maxwell's equations for electromagnetic fields. The \Schrodinger, vorticity equations, and wave equations follow from these equations as a natural outcome.
 
The article is organized as follows. 
Sec.~\ref{sec2} considers the continuity equation on 3D Euclidean space with three translations along axes $x, y, z$ and three rotations about these axes.
Here we introduce 6D space having six degrees of freedom due to three translations along axes $x, y, z$ and three  rotations about these axes. 
Here we go on from the SU(2) group to the quaternion group of $4\times4$ matrices for describing motions in 10D space (6+4 degrees of freedom) and introducing the balance equation in it.
In Sec.~\ref{sec3} we define the energy-momentum tensor and further based on the quaternion algebra we define differential shift operators by using four quaternion matrices. As a result of acting these operators on the energy-momentum tensor we get the force density tensor containing the irrotational force densities and orbital ones.
In Sec.~\ref{sec4} we get a set of the gravitomagnetic equations that are similar to Maxwell's equations of the electromagnetic field.
In Sec.~\ref{sec5} we derive from the gravitomagnetic equations three equations, namely: (subsec.~\ref{sec51})~the \Schrodinger equation (in particular, we consider here deceleration of a baryon-matter object in a cosmic space), (subsec.~\ref{sec52})~the vorticity equation (we consider here the neutron spin resonance experiment combined with the spin-echo spectroscopy for measuring gravitomagnetic effects), and (subsec.~\ref{sec53})~the wave equations describing gravitomagnetic waves in the superfluid quantum space-time.
In Sec.~\ref{sec6} we consider the cosmic microwave background radiation and its connection with the gravitomagnetic waves.
Sec.~\ref{sec7} summarizes the results.

\section{\label{sec2}Balance equation in 6D space}

Let us discuss the degrees of freedom of our 3D space.
Around each point $(x,y,z)$ of the space one can mentally describe the sphere of unit radius.
 It can mean that any rotation of the unit vector describes some path on the surface of the unit sphere
 with its origin fixed at the point $(x,y,z)$.
 
\begin{figure}[htb!]
  \begin{picture}(180,197)(0,0)
      \includegraphics[scale=0.45]{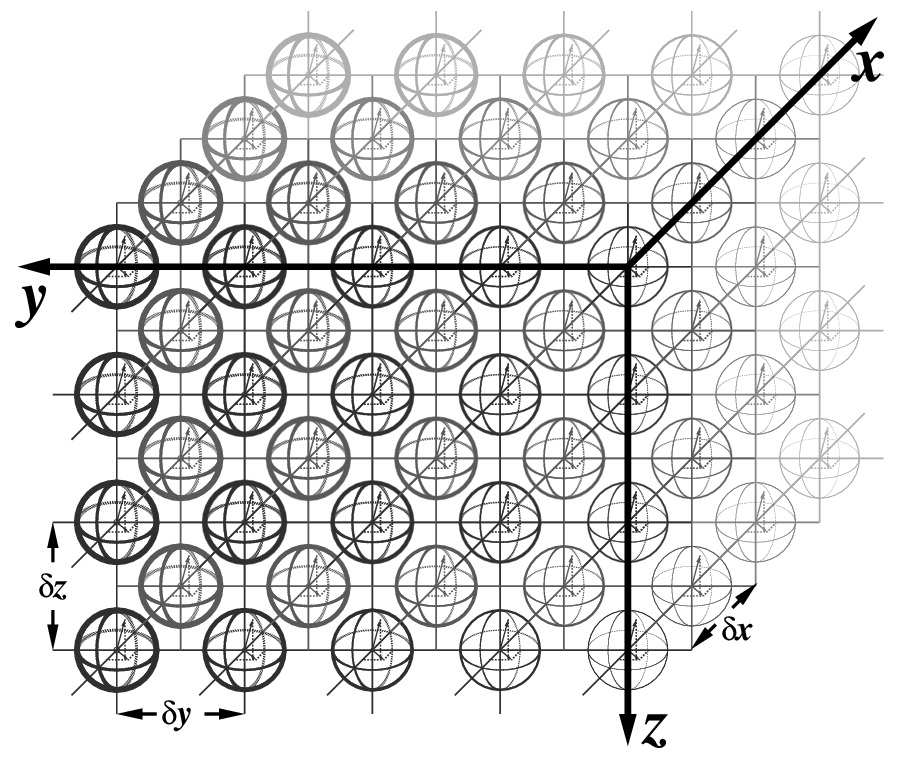}
  \end{picture}
  \caption{
 6D space $\Rc^6 = \Rc^{3}\!\otimes\!\Sc^{3}$ consists of 3D space $\Rc^3$, where around each point $(x,y,z)\in\Rc^3$ rotation of a gyro is allowed along any arc laying on 3D sphere $\Sc^{3}$. Distances $\delta x$, $\delta y$, $\delta z$ between these centers of the spheres
  tend to such minimal values as possible. 
 }
  \label{fig=1}
\end{figure}

 We know that a pseudovector is the mathematical image of a gyro.
 A gyro bundle containing the gyros with different inertial masses will show scattering the pseudovector tips on the unit sphere  surface at their rotations.
By combining the distribution of locations of the gyros in 3D space, $\Rc^{3}$, with the distribution of vertices of their pseudovectors on surface of 3D sphere, $\Sc^{3}$,  we come to the joint distribution of these rotating objects in 6D space $\Rc^6 = \Rc^{3}{\otimes}\Sc^{3}$,~Fig.~\ref{fig=1}.

 We begin first from  definition of a joint amplitude distribution of the translations and rotations on the space
 $\Rc^6 ={\mathcal R}^{3}\otimes{\mathcal S}^{3}$ that we record 
 as ${R\hspace{-6pt}R}({\vec r},t)=R({\vec r},t)|\varphi({\vec r},t)\rangle$.
 Here $R({\vec r},t)$ is the amplitude distribution of the translation flows in space
 ${\mathcal R}^{3}$ and $|\varphi({\vec r},t)\rangle$ is the amplitude distribution of rotation flows on the sphere ${\mathcal S}^{3}$.
 Note that translations and rotations are different forms of motion. The first is irrotational form, while the second is orbital. That's why we take a direct product of $R({\vec r},t)$ and $|\varphi({\vec r},t)\rangle$.

 Let us define a generator ${\mathcal D}({\vec{\mathit v}})$ that realizes
 a small shift of  ${R\hspace{-6pt}R}({\vec r},t)$ by $\delta\tau$ on the space ${\Rc}^{6}$:
\begin{eqnarray}
  {R\hspace{-6pt}R}({\vec r},t+{\delta\tau}) =
  {\mathcal D}({\vec{\mathit u}}){R\hspace{-6pt}R}({\vec r},t).
\label{eq=1}
\end{eqnarray}
 It is a balance equation.
 The generator ${\mathcal D}({\vec{\mathit u}})$ transforms the amplitude distribution of 
 the spin flows~\cite{AgamalyanEtAl1988,Sbitnev1989,Sbitnev2018c}, 
 but does not act on the amplitude distribution $R({\vec r},t)$.
 It looks as follows
\begin{equation} 
 {\mathcal D}({\vec{\mathit u}}) 
 =         u_{0}\sigma_{0}
 + {\bf i}u_{x}\sigma_{x} 
 + {\bf i}u_{y}\sigma_{y} 
 + {\bf i}u_{z}\sigma_{z} .
\label{eq=2}
\end{equation}
  Here ${\bf i}$  is the imaginary unit. 
  Four basic matrices in this expression, namely,
 three Pauli matrices, $\sigma_{x}$, $\sigma_{y}$, and $\sigma_{z}$, and the unit matrix $\sigma_{0}$ read:
\begin{equation}
  \sigma_{x} = 
  \left(
        \matrix{
                     0  &  ~~1 \cr
                     1  &  ~~0 \cr
                  }
  \right),~
  \sigma_{y} = 
  \left(
        \matrix{
                     0         &   -{\bf i} \cr
                     {\bf i}  &    ~~0    \cr
                  }
  \right),~
   \sigma_{z} = 
  \left(
        \matrix{
                     1  &  ~~0 \cr
                     0  &  -1    \cr
                  }
  \right),~
  \sigma_{0} = 
  \left(
        \matrix{
                     1  &  ~~0 \cr
                     0  &  ~~1 \cr
                  }
  \right).
\label{eq=3} 
\end{equation}
The matrices submit to the following commutation relations
\begin{eqnarray}
\nonumber
 &&\hskip-16pt
 \sigma_{x}\sigma_{y}=-\sigma_{y}\sigma_{x}={\bf i}\sigma_{z},~~~
 \sigma_{y}\sigma_{z}=-\sigma_{z}\sigma_{y}={\bf i}\sigma_{x},      
 \\
 &&\hskip-16pt
 \sigma_{z}\sigma_{x}=-\sigma_{x}\sigma_{z}={\bf i}\sigma_{y},~~~
 \sigma_{x}^{2} = \sigma_{y}^{2} = \sigma_{z}^{2} = \sigma_{0}
\label{eq=4}
\end{eqnarray}   
 As a result, the shifting transformation~(\ref{eq=1}) can be reduced to two equations:
\begin{eqnarray}
\label{eq=5}
 {{d}\over{d\,t}}\rho({\vec r},t) &=& 0,\\
 \varphi({\vec r},t+{\delta\tau})\rangle  
 &=& {\mathcal D}({\vec{\mathit u}}) \varphi({\vec r},t)\rangle .
\label{eq=6}
\end{eqnarray}
Here $\rho({\vec r},t) = R^2({\vec r},t)$ is the density distribution of sub-quantum particles, carriers of masses.
Therefore, Eq.~(\ref{eq=5}) is the continuity equation.
It says that there are neither sources nor sinks in the space ${\mathcal R}^{3}$. 
While Eq.~(\ref{eq=6}) implies  existence of external fields that perturb motions of spins~\cite{AgamalyanEtAl1988}.

The above transformation associates the translation on the space ${\mathcal R}^{3}$ with the rotations of the sphere ${\mathcal S}^{3}$ of unit radius with applying the unitary unimodular group SU(2)~\cite{Hamermesh1962,PenroseRindler1984}.
For a more complete creative perception of these transformations, it makes sense to go to  the quaternion representation of the motions  - to set of real $4\Times4$ matrices.

\subsection{\label{sec21}Quaternion representation}

The generators ${\mathcal D}({\vec{\mathit u}})$ and ${\mathcal D}({\vec s})$ belong to the group SU(2). Then their multiplication 
\begin{eqnarray}
\label{eq=7}
  &&  {\mathcal D}({\vec {\tilde s}}) =  {\mathcal D}({\vec{\mathit u}})\cdot{\mathcal D}({\vec s}) \;\Rightarrow\;
 ({\tilde s}_0\sigma_0 + {\bf i}{\tilde s}_x\sigma_x + {\bf i}{\tilde s}_y\sigma_y + {\bf i}{\tilde s}_z\sigma_z)
\\ \nonumber
&=& 
({u}_0\sigma_0 + {\bf i}{u}_x\sigma_x + {\bf i}{u}_y\sigma_y + {\bf i}{u}_z\sigma_z)\cdot
({s}_0\sigma_0 + {\bf i}{s}_x\sigma_x + {\bf i}{s}_y\sigma_y + {\bf i}{s}_z\sigma_z)
\end{eqnarray}
belongs to the group SU(2) as well.
By putting into account the multiplication order of the Pauli matrices~(\ref{eq=4})
we find representation of Eq.~(\ref{eq=7}) through 4$\times$4 matrix multiplication
\begin{equation}
\left(
       \matrix{
                  {\tilde s}_{0} \cr
                  {\tilde s}_{x} \cr
                  {\tilde s}_{y} \cr
                  {\tilde s}_{z} \cr
                 }
\right) =
\left(
      \matrix{
                 u_{0}   &   -u_{x}  &   -u_{y}  &   -u_{z}  \cr
                 u_{x}   &~~u_{0}  &~~u_{z}  &   -u_{y}  \cr
                 u_{y}   &   -u_{z}  &~~u_{0}  &~~u_{x}  \cr
                 u_{z}   &~~u_{y}  &   -u_{x}  &~~u_{0}  \cr
                }       
\right)
\left(
       \matrix{
                  s_{0} \cr
                  s_{x} \cr
                  s_{y} \cr
                  s_{z} \cr
                 }
\right)
\label{eq=8}
\end{equation} 
\begin{figure}[htb!]
\begin{picture}(180,120)(0,0)
      \includegraphics[scale=0.6]{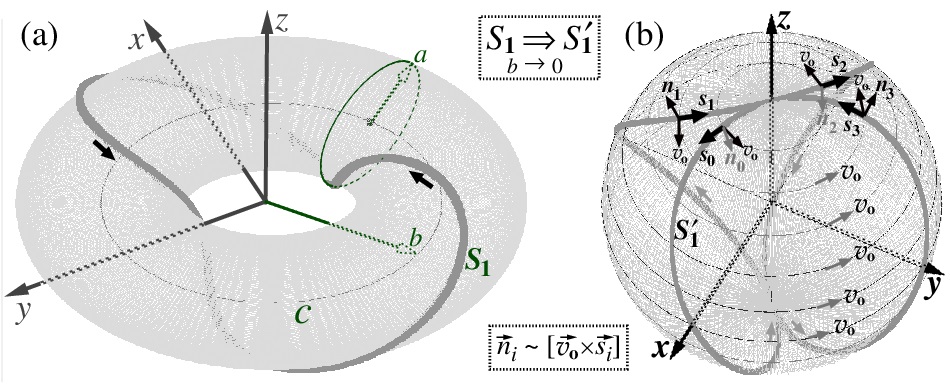}
  \end{picture}
  \caption{Representation of a spin rotation in 3D space~\cite{Sbitnev2016c,Sbitnev2018b,Sbitnev2018c}:
(a)~vortex tube of radius $a=1$ rolled into a torus having radius $b$. The torus is enveloped by string $S_1$, along which a current flows indicated by black arrows. 
At $b\rightarrow 0$ the string $S_1$ is mapped on the string $S'_1$ enveloping vortex ball, shown in figure from the left:
(b)~black arrows ${\vec{\mathit s}}_0$, ${\vec{\mathit s}}_1$, ${\vec{\mathit s}}_2$, ${\vec{\mathit s}}_3$ on this string show orientation of a flag rotating about this sphere along the string $S_1'$. 
The flag demonstrates rotation of spin-1/2, Fig.~\ref{fig=3}.
Orthogonal vectors ${\vec{\mathit n}}_i = [{\vec{\mathit v}}_0\Times{\vec{\mathit s}}_i]$, $i=0,1,2,3$, are oriented both outward (${\vec{\mathit n}}_1$, ${\vec{\mathit n}}_3$)  and inward (${\vec{\mathit n}}_0$, ${\vec{\mathit n}}_2$) of the sphere surface. That is, at moving the flag along the string, its orientation changes twice:
after the first revolution it flips  on 180 degrees. Only after the second revolution it reaches the full rotation on 360 degrees. 
 It is   manifestation of  the geometric phase~\cite{Berry1984,IoffeEtAl1991}. 
 }
  \label{fig=2}
\end{figure}
\begin{figure}[htb!]
  \begin{picture}(180,105)(0,0)
      \includegraphics[scale=0.33]{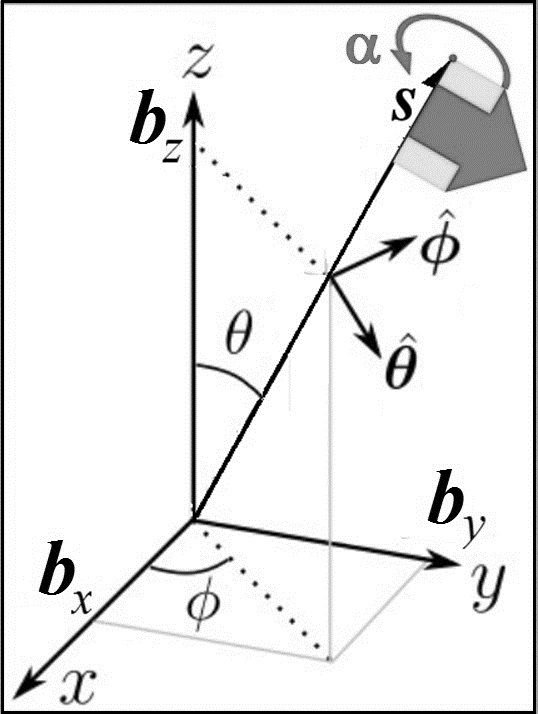}
  \end{picture}
  \caption{
Representation of a spin rotation in 3D space, see Fig.~\ref{fig=2}(b):
three dimensional unit pseudo-vector ${\vec{\mathit b}} = (b_x,b_y,b_z)$ is  endowed with an extra degree of freedom, by flag ${\vec{\mathit S}}$ that can  turn around  this vector by angle $\alpha$; 
 }
  \label{fig=3}
\end{figure}

One may extract from Eq.~(\ref{eq=8}) a set of four quaternion matrices~\cite{AgamalyanEtAl1988,Sbitnev2008a}:
\begin{eqnarray}
\nonumber
  &&
  \eta_{x} =
  \left(
        \matrix{
                    ~~0 &    -1 & ~~0 & ~~0 \cr
                    ~~1 & ~~0 & ~~0 & ~~0 \cr
                    ~~0 & ~~0 & ~~0 & ~~1 \cr
                    ~~0 & ~~0 &    -1 & ~~0 \cr
                   }
  \right),~
  \eta_{y} =
  \left(
        \matrix{
                    ~~0 & ~~0 &    -1 & ~~0 \cr
                    ~~0 & ~~0 & ~~0 &    -1 \cr
                    ~~1 & ~~0 & ~~0 & ~~0 \cr
                    ~~0 & ~~1 & ~~0 & ~~0 \cr
                   }
  \right),~  \\
  &&
  \eta_{z} =
  \left(
        \matrix{
                    ~~0 & ~~0 & ~~0 &    -1 \cr
                    ~~0 & ~~0 & ~~1 & ~~0 \cr
                    ~~0 &    -1 & ~~0 & ~~0 \cr
                    ~~1 & ~~0 & ~~0 & ~~0 \cr
                   }
  \right),~
  \eta_{0} =
  \left(
        \matrix{
                    ~~1 & ~~0 & ~~~0 & ~~0 \cr
                    ~~0 & ~~1 & ~~~0 & ~~0 \cr
                    ~~0 & ~~0 & ~~~1 & ~~0 \cr
                    ~~0 & ~~0 & ~~~0 & ~~1 \cr
                   }
  \right),~  
\label{eq=9}  
\end{eqnarray}
 which submit to the following rules of multiplication
\begin{eqnarray}
\nonumber
 &&
 \hskip-26pt
 \eta_{x}\eta_{y}=-\eta_{y}\eta_{x}=-\eta_{z},~~
 \eta_{y}\eta_{z}=-\eta_{z}\eta_{y}=-\eta_{x},      
 \\
 &&
 \hskip-26pt
 \eta_{z}\eta_{x}=-\eta_{x}\eta_{z}=-\eta_{y},~~ 
 \eta_{x}^{2} = \eta_{y}^{2} = \eta_{z}^{2} \;= -\eta_{0}.
\label{eq=10}
\end{eqnarray}   

One can expand the generator~(\ref{eq=2}) in the basis of the quaternion matrices
\begin{equation} 
 {\mathcal D}({\vec{\mathit u}}) 
 = u_{0}\eta_{0}
 + u_{x}\eta_{x} 
 + u_{y}\eta_{y} 
 + u_{z}\eta_{z} .
\label{eq=11}
\end{equation}
 And by rewriting Eq.~(\ref{eq=6}) we get the following equation
\begin{equation}
 |\varphi({\vec r},t+{\delta\tau})\rangle  =   {\mathcal D}({\vec{\mathit u}}) |\varphi({\vec r},t)\rangle.
\label{eq=12}
\end{equation} 
 The spinor $|\varphi({\vec r},t)\rangle$ represented by the four real variables $s_0$, $s_x$, $s_y$, and $s_z$ looks as follows
\begin{equation}
 |\varphi({\vec r},t)\rangle =
 \left(
    \matrix{
              \varphi_{\uparrow} \cr
              \varphi_{\downarrow}
              }
 \right)
 =
 \left(
    \matrix{
              ~s_0 + {\bf i}s_z \cr
              {\bf i}(s_x+{\bf i}s_y)
              }
 \right).
\label{eq=13}
\end{equation} 
 The variables  $s_x$, $s_y$, and $s_z$ mark the tip of  pseudo-vector on the sphere ${\mathcal S}^3$, whereas the variable $s_0$ shows orientation of the flag of this pseudo-vector moving along the string ${\mathit S}_1'$ , see Figs.~\ref{fig=2} and~\ref{fig=3}. 
 
Fig.~\ref{fig=2}(a) shows the torus enveloped by the string ${\mathit S}_1$ twisting round two times about the torus tube.
The topological mapping of the toroidal vortex to the vortex ball at tending the torus radius $b$ to zero translates the string ${\mathit S}_1$ to ${\mathit S}_1'$ that envelops the ball two times, see Fig.~\ref{fig=2}(b). The ball rotates about $z$ axis with the orbital velocity $\vec{\mathit v}_o$~\cite{Sbitnev2016c}. 
Arrows ${\mathit s}_i$, $i = 0,1,2,3$ drawn along the string ${\mathit S}_1'$ indicate orientations of a flag. It is the fourth degree of freedom prescribed to the three-dimensional pseudo-vector for describing the spin, see Fig.~\ref{fig=3}
Observe that the vector product of the orbital velocity $\vec{\mathit v}_o$ by the flag ${\mathit s}_i$ gives the vector ${\mathit n}_i$ perpendicular to the surface of the sphere. It can be oriented either inside of the sphere or out. In a whole, these vectors are oriented in opposite directions in the vicinity of each point lying on the sphere.  This double-layer surface differs from the Klein bottle~\cite{Rapoport2009,Rapoport2016a,Rapoport2016b}. Here  inversions of the orientations of  the unit vector ${\vec{\mathit n}}$ occur at infinitesimal shift along the strings passing through both opposite poles of the sphere.

In Fig.~\ref{fig=3} the flag is drawn as a big dark gray arrow on the background of light gray sheet. The flag represents an extra degree of freedom - it can rotate about the unit vector ${\vec b}=(b_x, b_y, b_z)$ on angle $\alpha$. The unit pseudo-vector ${\vec b}$ has two degrees of freedom - the tangential angle $\theta$ and the orbital angle $\phi$.  
 The added third degree of freedom, $\alpha$, translates motion of this pseudo-vector onto the 3D sphere of unit radius, what corresponds to the motion of spin.
 Observe that orientation of the flag demonstrates manifestation of the geometric phase at rotation of a spin on 3D sphere~\cite{Berry1984,IoffeEtAl1991}.
  Now we can define real coefficients $u_0$, $u_x$, $u_y$, $u_z$  of the quaternion matrix ${\mathcal D}({\vec u})$ in equations~(\ref{eq=11})-(\ref{eq=12}):
\begin{equation}
 u_{0} = \cos\biggl({{\alpha}\over{2}}\biggr),~~~
 u_{x} = b_{x}\sin\biggl({{\alpha}\over{2}}\biggr),~
 u_{y} = b_{y}\sin\biggl({{\alpha}\over{2}}\biggr),~
 u_{z} = b_{z}\sin\biggl({{\alpha}\over{2}}\biggr).
\label{eq=14}
\end{equation}

 \section{\label{sec3}Energy-momentum tensor  in the quaternion basis}   

Let us first define differential quaternion operators as follows 
\begin{equation}
\left.
   \matrix{
        {\mathcal D}\hskip6pt &\, ={\bf i}c^{-1}\partial_t\eta_0
              \:+ \partial_x\eta_x\: + \partial_y\eta_y\: + \partial_z\eta_z, \cr
        {\mathcal D}^{\bf T}\!& ={\bf i}c^{-1}\partial_t\eta_0^{\bf T}
              + \partial_x\eta_x^{\bf T} + \partial_y\eta_y^{\bf T} + \partial_z\eta_z^{\bf T} \cr
          &\, =   {\bf i}c^{-1}\partial_t\eta_0\:
              - \partial_x\eta_x\: - \partial_y\eta_y\: - \partial_z\eta_z.
              }
\right.
\label{eq=15}
\end{equation}
 Here $\partial_t =\partial/\partial t$, $\partial_x =\partial/\partial x$, etc., 
 $c$ is the speed of light, and sign ${\bf T}$ means the transposition\footnote[1]{in a Clifford bundle setting $\mathcal D$ would identify as a Dirac-Kahler operator~\cite{RodriguesAndCdeO2007}}.
 The d'Alembertian, wave operator, for the case of the negative metric signature
 ($g_{00}=-1$, $g_{11}=g_{22}=g_{33}=1$) 
 looks as follows
\begin{equation}
   {\mathcal D}^{\bf T}{\mathcal D} = {\mathcal D}{\mathcal D}^{\bf T}  = \
    (-c^{-2}\partial_t^2 + \partial_x^2 + \partial_y^2 + \partial_z^2)\eta_0 = \square.
\label{eq=16}
\end{equation}

 Let $\epsilon$ be an energy density, while $p_x$, $p_y$, $p_z$ are components of the momentum density.
 The latter multiplied by the speed of light, $c$, is adopted as a renormalized momentum density
\begin{equation} 
 {\vec p}\leftarrow{\vec p}\cdot c 
 =  \gamma({\mathit v})\rho_{M} {\vec{\mathit v}} 
     \Cdot c
      \approx \rho_{M} {\vec{\mathit v}}\!\cdot\! c.
\label{eq=17} 
\end{equation}
In turn, the relativistic energy density, $\epsilon$ can be expressed in terms of the momentum density by the following expression 
\begin{equation} 
  \epsilon = \sqrt{ \rho_{M}^2c^4 + p^2} = \gamma({\mathit v}) \rho_{M} c^2
  \approx \rho_{M} c^2 + \rho_{M}{{\mathit v^2}\over{2}}.
 \label{eq=18}
 \end{equation}
 Here $\rho_{M} = \rho m$ is the mass density of a carrier contained in the unit volume~${\Delta V}$
 and $\gamma({\mathit v})=(1-{\mathit v}^2/c^2)^{-1/2}$ is the Lorentz factor.
Both values, $\epsilon$ and ${\vec p}$, are seen to have the same dimensionality of the energy density~\cite{Sbitnev2015c}, which is the dimensionality of pressure,  $\rm Pa = kg\cdot\!m^{-1}\cdot\!s^{-2}$.
 
 Observe that the values $\epsilon$ and ${\vec p}$ are invariant with respect to the gauge transformation
\begin{equation}
\left\{
 \matrix{
          \epsilon & \Rightarrow & \epsilon \;-\; c^{-1}\partial_t\phi,  \cr
          {\vec p} & \Rightarrow & {\vec p} \;+\; \nabla\phi,~~~~
        }
\right.
\label{eq=19}
\end{equation}
 here $\phi$ is an arbitrary scalar field, having dimensionality of  Energy$\times$Length$^{-2}$.
By combining the algebras of the quaternions~\cite{MarcerRowlands2017} with the energy and momentum terms written above
 we write out the energy-momentum density tensor with added the extra term ${\mathcal D}^{\bf T} \phi$:
\begin{equation}
               T = {\bf i}\epsilon\eta_0 + p_x\eta_x + p_y\eta_y + p_z\eta_z - {\mathcal D}^{\bf T} \phi.
\label{eq=20}
\end{equation}
The term ${\mathcal D}^{\bf T}\phi$ returns the expression 
${\bf i}c^{-1}\partial_t\eta_0\phi-(\nabla\Cdot{\vec\eta})\phi$ 
that comes from  the arbitrary scalar field $\phi$  added in Eq.~(\ref{eq=19}).
 Further we introduce the Lorentz gauge condition
\begin{equation}
 {{1}\over{4}}{\rm trace}\;{\mathcal D}{T} =-{{1}\over{c}}\partial_t\epsilon
 - \partial_xp_x  - \partial_yp_y  - \partial_zp_z  - \square\phi = 0 
\label{eq=21}
\end{equation}  
 It describes the general energy-momentum conservation law becuase of presence of the arbitrary scalar field $\phi$.
 Here the term $\square\phi = {\mathcal D}^{\bf T} {\mathcal D}\phi$ represents the wave equation for the scalar field $\phi$.

  Let us now write out  the product ${\mathcal D}\!\cdot\!{T}$ in details
\begin{eqnarray}
\nonumber
 {\mathcal D}\!\cdot\!{T} &=& 
 \biggl(
   -{{1}\over{c}}{{\partial\epsilon}\over{\partial t}}
   - {{\partial p_x}\over{\partial x}}
   - {{\partial p_y}\over{\partial y}}
   - {{\partial p_z}\over{\partial z}} - \square\phi
 \biggr)\eta_0
\\ \nonumber
 &+&
 \biggl(
   {\bf i}
   \biggl(
     {{\partial\epsilon}\over{\partial x}}+{{\partial p_x}\over{c\partial t}}
   \biggr)
- \biggl(
   {{\partial p_z}\over{\partial y}} - {{\partial p_y}\over{\partial z}}
  \biggr)
 \biggr)\eta_x 
\\ \nonumber
 &+&
 \biggl(
   {\bf i}
   \biggl(
     {{\partial\epsilon}\over{\partial y}}+{{\partial p_y}\over{c\partial t}}
   \biggr)
- \biggl(
   {{\partial p_x}\over{\partial z}} - {{\partial p_z}\over{\partial x}}
  \biggr)
 \biggr)\eta_y
\\ 
 &+&
 \biggl(
   {\bf i}
   \biggl(
     {{\partial\epsilon}\over{\partial z}}+{{\partial p_z}\over{c\partial t}}
   \biggr)
- \biggl(
   {{\partial p_y}\over{\partial x}} - {{\partial p_x}\over{\partial y}}
  \biggr)
 \biggr)\eta_z 
\label{eq=22}
\end{eqnarray}
 The expression at the quaternion $\eta_0$ represents the Lorentz gauge when it is zero.
 Only the expressions at quaternions $\eta_x$, $\eta_y$, $\eta_z$ remain, which do not contain terms with $\phi$.
 The latter cancel each other.

  According to the equations given in~(\ref{eq=19}), the above expressions represent $x$, $y$, $z$
 components of 
 gavitoelectric and gyromagnetic fields\footnote[2]{Here we distinguish two components of the gravitomagnetic field - gravitoelectric  and gyromagnetic fields. It is due to that masses attract each other like opposite electric charges, while  gyroscopes create the vorticity like  magnets creating the magnetic induction.}.
 So, the above equation represents the force density tensor
\begin{equation}
 {\mathbb F}_{\Xi\Omega} = {\mathcal D}\!\cdot\!{T} =
\left(
  \matrix{
                       0     & ~~\Omega_x - {\bf i}\Xi_x  & ~~\Omega_y - {\bf i}\Xi_y   & ~~\Omega_z - {\bf i}\Xi_z \cr
    -\Omega_x + {\bf i}\Xi_x  & 0                           &    -\Omega_z + {\bf i}\Xi_z  & ~~\Omega_y - {\bf i}\Xi_y \cr
    -\Omega_y + {\bf i}\Xi_y  &~~\Omega_z -{\bf i}\Xi_z    &      0                       &    -\Omega_x +{\bf i}\Xi_x \cr
    -\Omega_z + {\bf i}\Xi_z  &   -\Omega_y+{\bf i}\Xi_y     & ~~\Omega_x - {\bf i}\Xi_x   &     0                     \cr
            }
\right) .
\label{eq=23}
\end{equation}
 At computation of derivatives of the energy density and the momentum density by time and by space these terms contain the functions $\rho({\vec r},t)$, $\gamma({\mathit v})$ containing ${\mathit v}^{2}({\vec r},t)$, and the velocity ${\vec{\mathit v}}({\vec r},t)$, see Eqs.~(\ref{eq=17}) and~(\ref{eq=18}). All these terms are subjected to computation of derivatives by time and length. 
 
 For simplicity we will consider further the non-relativistic limit.
In this case we omit the
 computation of derivatives of the Lorentz factor $\gamma({\mathit v})$.
 Results of such truncated computations give the following vectors 
\begin{eqnarray}
\label{eq=24}
  {\vec\Omega} &=& [\nabla\!\times\!{\vec p}] = 
   c [\nabla\!\times\!\rho_{M}{\vec{\mathit v}}] ,\\
  {\vec\Xi} &=& {{\partial{\vec p}}\over{c\partial t}} + \nabla \epsilon. 
\label{eq=25}
\end{eqnarray} 
Their dimensions are the force per volume, ${\rm N}\Cdot{\rm m}^{-3}$.
Observe that the vectors $\vec{\Omega}$ and  $\vec{\Xi}$ are similar to the magnetic and electric fields, $\vec B$ and $\vec E$, if we change the sign in Eq.~(\ref{eq=25}) and consider that $\vec p$ is the vector potential, $\vec A$, and $\epsilon$ is the scalar potential, $\phi$: ${\vec E}=  -c^{-1}\partial_t{\vec A} -\nabla{\phi}$.  However, change of sign at $\vec{\Xi}$ is not a good idea. Further we will see, that the vector $\vec{\Xi}$ will represent the force density in the Navier-Stokes equation which enters with positive sign\footnote[3]{It should be noted that the Navier-Stokes equation is the case of a trace-torsion geometry given by the velocity $\vec{\mathit v}$~\cite{Rapoport2000,Rapoport2003,Rapoport2005}. }.

The term  $ [\nabla\!\times\!\rho_{M}{\vec{\mathit v}}]$ in equation~(\ref{eq=24}) is decomposed into the sum of two terms $\rho_{M}[\nabla\!\times\!{\vec{\mathit v}}]$ and 
$[(\nabla\!\cdot\!\rho_{M})\!\times\!{\vec{\mathit v}}]$
The first term represents the vorticity ${\vec\omega}=[\nabla\!\times\!{\vec{\mathit v}}]$. 
The angular moment density ${\vec\Omega}$, proportional to the vorticity ${\vec\omega}$, depends on the mass distribution in the rotating body.
This term,  likewise the magnetic induction ${\vec B}$,  induces  precession of the massive body about direction of this moment. Therefore, returning to the unit vector ${\vec b}$ and the precession angle $\alpha$ shown in Eq.~(\ref{eq=14}) these quantities can be expressed as follows through the angular moment density
\begin{equation}
  {\vec b} = {{\vec\Omega}\over{\sqrt{\Omega_{x}^2+\Omega_{y}^2+\Omega_{z}^2}}}
\label{eq=26}
\end{equation} 
 and
\begin{equation}
  \alpha = -\gamma_{m}\sqrt{\Omega_{x}^2+\Omega_{y}^2+\Omega_{z}^2}\cdot{\delta\tau}.
\label{eq=27}
\end{equation} 
 Here $\gamma_{m}\sim 1/(\rho_{M} c)$ is the parameter due to which the angular momentum is reduced to the vorticity, ${\vec\omega}={\vec\Omega}/(\rho_{M}c)$.
 In some way this parameter is analogous to the gyromagnetic ratio, which is the ratio of its magnetic moment
 to the angular momentum~\cite{MartinsPinheiro2008,MartinsPinheiro2009,Martins2012}.
 While, the second term, $[(\nabla\!\cdot\!\rho_{M})\!\times\!{\vec{\mathit v}}]$, can be easily recognized if we remember the rotation of a raw egg.  
 Its inner fluid contents experiencing permanent deformation inhibits the rotation. 
 

Some remarks should be done relating to inner components of the velocity. 
First note that the velocity can consists of the orbital component (also named  solenoidal component) and the irrotational component coming from the gradient of the scalar field $\Phi$, namely, 
\begin{equation}
{\vec{\mathit v}} = -\nabla\Phi +[\nabla\Times{\vec A}].
\label{eq=28}
\end{equation}
The scalar and vector fields, $\Phi$ and $\vec A$\footnote[4]{Observe that  $\Phi$, having dimension [length$^2$/time],  is equal to $S/m$, where $S$ is the action function and $m$ is  mass, and ${\vec A}$ is proportional to the angular momentum divided by mass.},
follow from the Helmholtz-Hodge decomposition~\cite{Rapoport1997,Rapoport2005a,Rapoport2005} for the velocity ${\vec{\mathit v}}$:
\begin{equation}
   \Phi(\vec r) = {{1}\over{4\pi}}\int\limits_V
                       {{({\nabla}'\Cdot\vec{\mathit v}({\vec r}'))}\over{|\vec r - {\vec r}'|}} dV'
                    - \underbrace{{{1}\over{4\pi}}\oint\limits_S
                       {{({\vec n}'\Cdot\vec{\mathit v}({\vec r}'))}\over{|\vec r - {\vec r}'|}} dS'}_{{\rm for}~V={\mathcal R}^3~{\rm it~vanishes}}
\label{eq=29}
\end{equation}
and
\begin{equation}
  {\vec A}(\vec r) = {{1}\over{4\pi}}\int\limits_V
                       {{[{\nabla}'\Times\vec{\mathit v}({\vec r}')]}\over{|\vec r - {\vec r}'|}} dV'
                    - \underbrace{{{1}\over{4\pi}}\oint\limits_S
                       {{[{\vec n}'\Times\vec{\mathit v}({\vec r}')]}\over{|\vec r - {\vec r}'|}} dS'}_{{\rm for}~V={\mathcal R}^3~{\rm it~vanishes}}
\label{eq=30}
\end{equation}
Here $V\in{\mathcal R}^3$ and $S$ is the surface that encloses the domain $V$.

Note that the operator curl applied to the velocity ${\vec{\mathit v}}$ gives the pseudo-vector ${\vec\omega}=[\nabla\Times{\vec{\mathit v}}]$, named vorticity. 
According to the Stokes's theorem~\cite{KunduCohen2002}, the circulation
\begin{equation}
  {\mit\Gamma} = \oint\limits_{\partial S} {\vec{\mathit v}}\, d{\vec{\mathit l}}
  = \int\limits_S\hskip-6pt\int {\vec{\omega}}\,d{\mathbf S},
\label{eq=31}
\end{equation}
can be non zero. Here $\partial S$ means the boundary of a closed surface $S$.
The circulation returns the vorticity  
${\vec\omega}=[\nabla\!\times\!{\vec{\mathit v}}] = [\nabla\!\times\!{\vec{\mathit v}}_{\,\rm o}]$.
Here we represent the velocity 
${\vec{\mathit v}}$  by sum of two addends ${\vec{\mathit v}}_{\,\rm s}$ and ${\vec{\mathit v}}_{\,\rm o}$.  
The subscript o relates to an orbital velocity ${\vec{\mathit v}}_{\,\rm o}$
that obeys to $ [\nabla\!\times\!{\vec{\mathit v}}_{\,\rm o}]\ne 0$ and $(\nabla\!\cdot\!{\vec{\mathit v}}_{\,\rm o}) = 0$.
This velocity is named in literature~\cite{Lighthill1986} the solenoidal velocity. In our previous works this subscript was marked by the letter $R$ meaning the rotational velocity. Now, instead of the subscript $R$ we decided here to write the subscript o. 
This sign associates with the circulation around the closed loop as written in the circulation integral~(\ref{eq=31}). 
The subscript s points to existence of a scalar field $S$. The velocity ${\vec{\mathit v}}_{\,\rm s}$ is proportional to gradient of this field, 
$S$ means the action function. 
The velocity ${\vec{\mathit v}}_{\,\rm s}$ is named the irrotational velocity~\cite{Lighthill1986}.
It means that 
$[\nabla\!\times\!{\vec{\mathit v}}_{\,\rm s}] = 0$. 
So, the velocities ${\vec{\mathit v}}_{\,\rm s}$ and ${\vec{\mathit v}}_{\,\rm o}$ satisfy the following equations
\begin{equation}
\label{eq=32}
 \left.\hskip-12pt
    \matrix{
 (\nabla\cdot{\vec{\mathit v}}_{\,\rm o})  =  0~(\rm a);\hskip24pt&[\nabla\times{\vec{\mathit v}}_{\,\rm o}]={\vec\omega}~~(\rm b); \cr
 (\nabla\cdot{\vec{\mathit v}}_{\,\rm s}) \ne 0~(\rm c);\hskip24pt&\,[\nabla\times{\vec{\mathit v}}_{\,\rm s}]=0~~~(\rm d). \cr
           }
\hskip-12pt           
 \right.
\end{equation}
One can see that these formulas directly follow from Eq.~(\ref{eq=28}) supplemented by the Helmholtz-Hodge decompositions stated above.

\section{\label{sec4}Gravitomagnetic equations}

The function $\vec\Xi$ in Eq.~(\ref{eq=25}) represents a convective force acting on the fluid elements.
In the non-relativistic limit it reduces to the Navier-Stokes equation 
\begin{equation}
{\vec\Xi}={{\partial{\vec p}}\over{c\partial t}} + \nabla \epsilon = {\vec F},
\label{eq=33}
\end{equation}
where modified right part  ${\vec{F}}$ contains the external and internal force densities~\cite{Sbitnev2018b}. 
The external force density is represented through gradient of the potential energy $U$, as, for example, the gravitational potential energy of an object that depends on its mass and its distance from the centers of masses of another objects. 
The internal force density results from  gradient of the quantum potential $Q$ following from  the modified internal pressure gradient~\cite{Sbitnev2016b}. 
Also an extra term is added, that takes into account, in the general case, the viscous stress tensor of a fluid medium, fluctuating about zero~\cite{Sbitnev2015c}. 
When we try to justify the appearance of superfluidity this tensor reduces to the noise source tensor~\cite{Rapoport2007}. However, in order to avoid complicated computations further we will deal with only a scalar noise source~\cite{Sbitnev2016b}.

Repeating the expressions which underlie Maxwell's electromagnetic theory we define the density distribution of all acting on the fluid body forces
\begin{equation}
 \wp = {{1}\over{4\pi}}
 (\nabla\Cdot{\vec{F}}).
\label{eq=34}
\end{equation}
Let us now define 3D current density ${\vec \Im}={\vec{\mathit v}}\wp$. 
Further we define the 4D current density as follows
\begin{equation}
 {\vec {\bJ}} = {\bf i}c\wp\eta_0 + \Im_x\eta_x + \Im_y\eta_y + \Im_z\eta_z.
\label{eq=35}
\end{equation}
 The continuity equation in this case takes the following view
\begin{equation}
 {{1}\over{4}}{\rm trace}\;{\mathcal D}\,{\vec {\bJ}} =
 -{\partial_t\wp} - {\partial_x \Im_x} - {\partial_y \Im_y} - {\partial_z \Im_z} = 0,
\label{eq=36}
\end{equation} 
 that can be rewritten in a more evident form
\begin{equation}
   {{\partial\wp}\over{\partial t}} + (\nabla\!\cdot\!{\vec \Im}) = 0.
\label{eq=37}
\end{equation} 
 In some sense, this equation is a manifestation of Newton's third law, the action-reaction law. 
The law says, that all forces acting on the fluid element are balanced among themselves.

\begin{figure*}[htb!]
  \begin{picture}(300,85)(0,0)
      \includegraphics[scale=0.5]{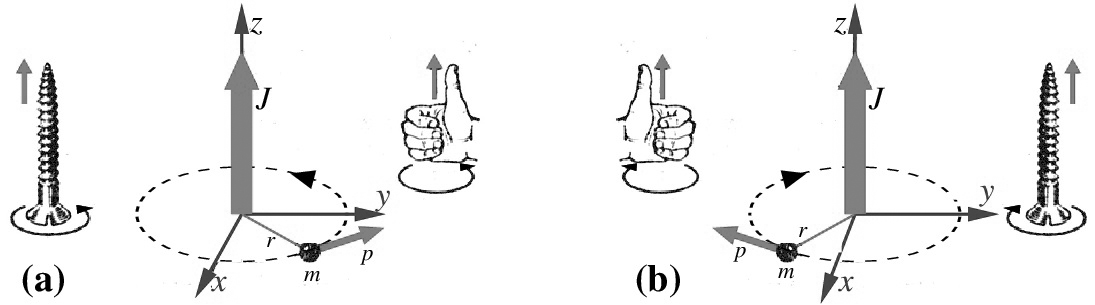}
  \end{picture}
  \caption{
 The angular momentum  ${\vec J}=[{\vec r}\Times{\vec p}]$ oriented by the rule of the right screw (a),
 and reflected in the plane $(x,z)$ (the mirror plane) now submits to the rule of the left screw (b).
 }
  \label{fig=4}
\end{figure*}  
  Let us now apply to the  force density tensor~(\ref{eq=23}) the differential operator ${\mathcal D}^{\bf T}$ 
  and equate it to the 4D current density ${\vec {\bJ}}$. We get the generating equation:
\begin{equation}
 {\mathcal D}^{\bf T}\!\cdot\!{\mathbb F}_{\Xi\Omega} = {{4\pi}\over{c}} {\vec {\bJ}},
\label{eq=38}
\end{equation}
 which in a Clifford bundle setting  is the Maxwell equation~\cite{Rapoport2005a,RodriguesAndCdeO2007}.
 By computing the product ${\mathcal D}^{\bf T}\!\cdot\!{\mathbb F}_{\Xi\Omega}$,  we obtain the following set of terms represented  as coefficients of the quaternion matrices $\eta_{0}$, $\eta_{x}$, $\eta_{y}$, $\eta_{z}$:
\begin{eqnarray}
\nonumber
 {\mathcal D}^{\bf T}\!\cdot\!{\mathbb F}_{\Xi\Omega} &=&
\{      - ({\partial_x \Omega_x} + {\partial_y \Omega_y} + {\partial_z \Omega_z})
+{\bf i}({\partial_x  \Xi_x} +  {\partial_y  \Xi_y} +  {\partial_z  \Xi_z})
\}\eta_0  
\\ \nonumber
  &+& 
\biggl\{\!\!
  \biggl(
    -{{1}\over{c}}{\partial_t \Xi_x} - ({\partial_y \Omega_z} - {\partial_z \Omega_y})
  \biggr)
 +{\bf i}  \biggl(
    -{{1}\over{c}}{\partial_t \Omega_x} + ({\partial_y \Xi_z} - {\partial_z \Xi_y})
  \biggr)
\!\!\biggr\}\eta_x
\\ \nonumber 
&+& 
\biggl\{\!\!
  \biggl(
   - {{1}\over{c}}{\partial_t \Xi_y} - ({\partial_z \Omega_x} - {\partial_x \Omega_z})
  \biggr)
 +{\bf i}  \biggl(
    -{{1}\over{c}}{\partial_t \Omega_y} + ({\partial_z \Xi_x} - {\partial_x \Xi_z})
  \biggr)
\!\!\biggr\}\eta_y
\\ \nonumber &+& 
\biggl\{\!\!
  \biggl(
  -  {{1}\over{c}}{\partial_t \Xi_z} - ({\partial_x \Omega_y} - {\partial_y \Omega_x})
  \biggr)
 +{\bf i}  \biggl(
    -{{1}\over{c}}{\partial_t \Omega_z} + ({\partial_x \Xi_y} - {\partial_y \Xi_x})
  \biggr)
\!\!\biggr\}\eta_z
\\  &=&
{{4\pi}\over{c}}({\bf i}c\wp\eta_0 + \Im_x\eta_x + \Im_y\eta_y + \Im_z\eta_z).
\label{eq=39}
\end{eqnarray}
 From here we get the following pairs of equations
\begin{eqnarray}
     && 
\label{eq=40}
    (\nabla\!\cdot\!{\vec \Omega}) = 0, \\ 
     &&    {\displaystyle
                 [\nabla\times{\vec \Xi}] - {{1}\over{c}}{{\partial}\over{\partial t}}{\vec \Omega}
              } = 0, 
\label{eq=41} 
\\&& 
\label{eq=42}
   (\nabla\!\cdot\!{\vec \Xi}) = 4\pi\wp, \\
      && 
         {\displaystyle
              [\nabla\times{\vec \Omega}] + {{1}\over{c}}{{\partial}\over{\partial t}}{\vec \Xi} = -{{4\pi}\over{c}}{\vec \Im}.  
          } 
\label{eq=43}
\end{eqnarray}
Dimension of these formulas is the same as $\wp$, namely ${\rm[ kg\cdot m^{-3}\cdot s^{-2}]}$.
It is the pressure per the area unit, ${\rm Pa\Cdot m^{-2}}$.

Eq.~(\ref{eq=40}) represented by the scalar product of the operator $\nabla$ by the vector $\vec\Omega$ defines a type of this vector  according to formulas~(\ref{eq=32}). It states that the force density vector $\vec\Omega$ is no irrotational but it relates to orbital motions.
 
 The second equation, Eq.~(\ref{eq=41}), results from the direct differentiation of Eqs.~(\ref{eq=24}) and~(\ref{eq=25}):
\begin{eqnarray}
\label{eq=44}
  {{\partial}\over{c\partial t}} {\vec\Omega} &=& {{\partial}\over{c\partial t}}[\nabla\Times{\vec p}], \\
  {[ \nabla\Times{\vec \Xi}]}
   &=&   [\nabla\Times  {{\partial}\over{c\partial t}}{\vec p}]
 + \underbrace{[\nabla\Times\nabla\epsilon]
                      }_{= 0}.
\label{eq=45}
\end{eqnarray} 
 Equality of these two expressions says that the vorticity once originated in the fluid keeps its chirality in time. It follows from the Kelvin's circulation theorem, that in a barotropic ideal fluid with conservative forces acting to the fluid, the circulation around a closed curve moving with the fluid remains constant with time~\cite{KunduCohen2002}.
 Observe that the circulation of the vector field under consideration,~(\ref{eq=34})-(\ref{eq=39}),  submits to the rule of the right screw, Fig.~\ref{fig=4}(a).

The third equation, Eq.~(\ref{eq=42}), in pair with the continuity equation~(\ref{eq=5}) can be reduced to the \Schrodinger equation  after defining the wave function in the polar form, $\psi = \sqrt{\rho}\Cdot\exp\{{\bf i}S/\hbar\}$.  It will be shown later.

It turns out, that the equations~(\ref{eq=40}) to~(\ref{eq=43}) are known in the scientific literature as the 
gravitomagnetic equations~\cite{Heaviside1893,CiufoliniWheeler1995,MashhoonEtAl2001,Ruggiero2016}.
They are similar to the Maxwell's equations:
\begin{eqnarray}
     && 
\label{eq=46}
    (\nabla\!\cdot\!{\vec B}) = 0, 
    \\ 
     &&      {\displaystyle
             [\nabla\times{\vec E}] + {{1}\over{c}}{{\partial}\over{\partial t}}{\vec B}
           } = 0 
\label{eq=47} 
\\&& 
\label{eq=48}
           (\nabla\!\cdot\!{\vec E}) = 4\pi\varrho, 
      \\
      && 
    {\displaystyle
            [\nabla\times{\vec B}] - {{1}\over{c}}{{\partial}\over{\partial t}}{\vec E} = {{4\pi}\over{c}}{\vec j}.  
    } 
\label{eq=49}
\end{eqnarray}
The exceptions relate to signs of Eqs.~(\ref{eq=41}) and~(\ref{eq=43}) that differ from  those of  Eqs.~(\ref{eq=47}) and~(\ref{eq=49}).
Accordance in signs of gravitomagnetic equations and Maxwell's equations can be achieved by reflecting  in plane $(x,z)$ (the mirror plane) the rotation of the massive body about $z$-axis, see Fig.~\ref{fig=4}(b).
 This operation is equivalent to  change of the rule from the right screw to the left screw. 
It is achieved by change of sign of  $\vec{\Omega}$ in Eqs~(\ref{eq=41}) and~(\ref{eq=43}). 

Note that Eqs.~(\ref{eq=46})-(\ref{eq=49})   were obtained by the same method as Eqs.~(\ref{eq=40})-(\ref{eq=43}). 
Namely, the electromagnetic tensor, $F_{\rm EM}$, has the same view as the tensor ${\mathbb F}_{\Xi\Omega}$ when we replace $\vec{\Xi}$ by $\vec E$ and $\vec{\Omega}$ by $\vec B$~\cite{Girard2007,Sbitnev2018c}. Exception is that for getting the Maxwell's equations~(\ref{eq=46})-(\ref{eq=49}) we apply the operator ${\mathcal D}$ to the tensor $F_{\rm EM}$.
Here  at differentiating~(\ref{eq=39}) we chose the operator ${\mathcal D}^{\bf T}$ with the aim to get the true equality of $c^{-1}\partial_t {\vec\Omega}$ and $[\nabla\Times\vec{\Xi}]$, as shown in~(\ref{eq=44}) and~(\ref{eq=45}).
\begin{figure}[htb!]
  \begin{picture}(170,120)(0,0)
      \includegraphics[scale=0.48]{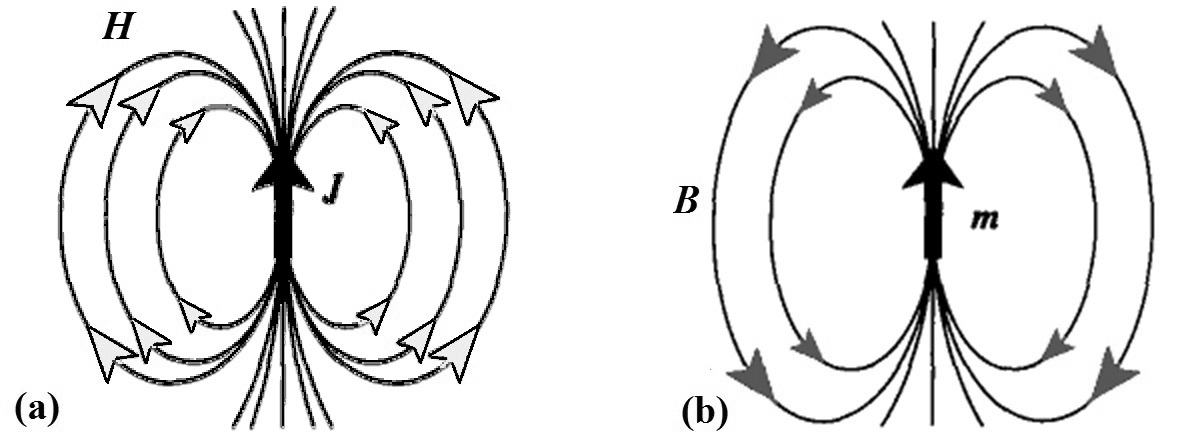}
  \end{picture}
  \caption{
 The directionalities of gravitomagnetism and magnetism compared and contrasted~\cite{CiufoliniWheeler1995}: 
(a) The gravitomagnetic field $\vec H$ in the weak field approximation, $\vec J$ is the angular momentum of the central body; 
(b) The magnetic induction $\vec B$ in the neighborhood of a magnetic dipole moment $\vec m$.
 }
  \label{fig=5}
\end{figure}  

\begin{figure}[htb!]
\begin{picture}(170,100)(0,0)
      \includegraphics[scale=0.35]{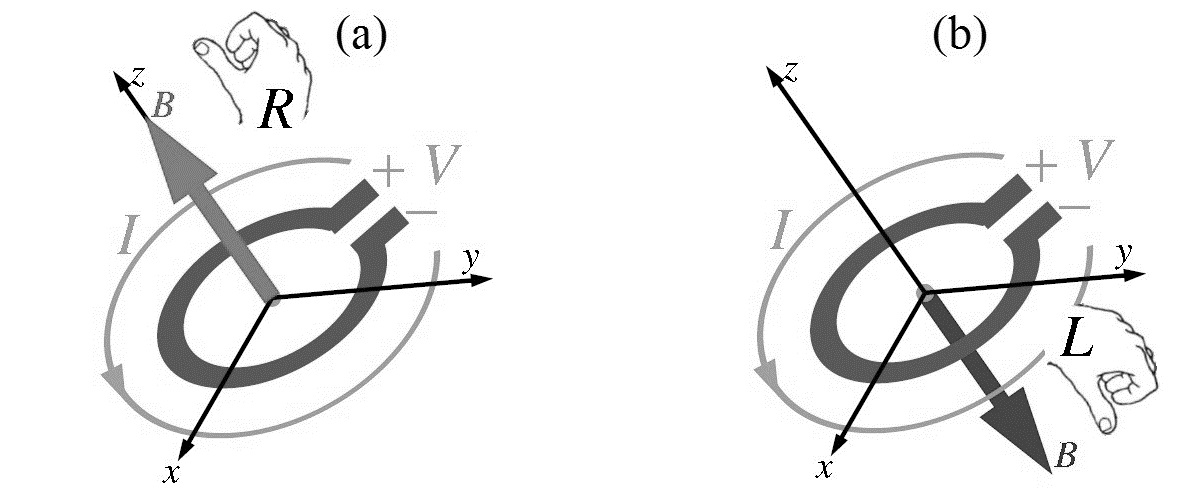}
  \end{picture}
  \caption{
 The law of chirality: right hand law (a) and left hand law (b) 
 for circulation of current in a ring with orientation of the magnetic induction.
 }
  \label{fig=6}
\end{figure}  

There is a some symmetry breaking between the gyro motion and the Maxwell's electromagnetic field. The first obvious breaking is due to different behavior of the masses and charges - the masses always attract each other but the same-name charges, as well as the same name poles of the magnets are repelled. 
This difference is well shown in Fig.~\ref{fig=5} taken from the book of Ignazio Ciufolini and John Archibald Wheeler entitled "gravitation and inertia"~\cite{CiufoliniWheeler1995}.
In Figures~(a) and~(b) orientations of the gravitomagnetic field and the magnetic induction are compared at equal orientations of the angular momentum~${\vec J}$  and the magnetic dipole momentum~${\vec m}$. The orientations of these two fields are opposite. The authors have depicted  this fact  by drawing oppositely oriented arrows of these force lines. In these figures the arrows, pointing opposite orientations of the force lines, are represented by different colors. The first are shown in white and the second in black.
We note that the authors  in their figure have also shown comparison  with a fluid drag~\cite{CiufoliniWheeler1995}. It should be noted, that many authors carry out deep analogies of turbulent, vortex flows of fluids  with the  electromagnetism by basing on a fluidic viewpoint~\cite{Kambe2010,Marmanis1998,Martins2012,MartinsPinheiro2008,MartinsPinheiro2009}.

The equations of gravitomagnetic and electromagnetic fields are almost identical. They allow a procedure of the one-to-one mapping to each other. For that, it is sufficient to change the sign in Eqs.~(\ref{eq=40})-(\ref{eq=43}) or of the variable $\vec{\Omega}$ or the both variables $\vec{\Xi}$ and $\wp$. And vice versa,  by default one can adopt the opposite orientation of the magnetic induction,  as shown in Fig.~\ref{fig=6}.

The equations of the gravitomagnetic fields are widely represented in the literature beginning from the earlier publication of Heaviside~\cite{Heaviside1893} up to nowadays~\cite{ClarkTucker2000,MashhoonEtAl2001,MatosTajmar2006,Ruggiero2016,Sekine2016,UmmarinoGallerati2017,BeheraNaik2004,Behera2017,CashenEtAl2017,SarkarEtAl2018}.
 
\subsection{\label{sec41}Quadratic forms of the gyromagnetic field tensor}

Two quadratic forms directly follow out from the force density tensor~(\ref{eq=23}).

\begin{itemize}
\item[$\bullet$]
The first quadratic form reads
\begin{equation}
{{1}\over{2}} {\mathbb F}_{\Xi\Omega}^{\,\dagger}{\mathbb F}_{\Xi\Omega}
= -W_0\eta_0 - {\bf i}W_x\eta_x - {\bf i}W_y\eta_y - {\bf i}W_z\eta_z.
\label{eq=50}
\end{equation}
 Here $\dagger$ is the sign of complex conjugation, 
$W_0 = (\Xi^2+\Omega^2)/2$, and ${\vec W}=[{\vec \Xi}\times{\vec \Omega}]$.
These terms are similar to the electromagnetic energy density, $(\varepsilon_0E^2+\mu_0^{-1}B^2)/2$, and the energy flux $\mu_0^{-1}[{\vec E}\Times{\vec B}]$. 
Dimensions of these values in SI are Joule/m$^3$ and Watt/m$^2 = {\rm (J/m^3)\Cdot(m/s)}$, respectively.
Here the vacuum permittivity $\varepsilon_0\approx8.85\Times10^{-12}~{\rm F\Cdot m^{-1}}$ and the permeability $\mu_0 = 4\pi\Times10^{-7}~{\rm H\Cdot m^{-1}}$.
In our case, however, dimension of both $\Xi^2$ and $\Omega^2$ is $\rm Pa^2\Cdot m^{-2}$.
To lead the dimensions of $W_0$ and $\vec W$ to those of the energy per volume and watt (a unit of power) per square we should introduce constants providing such links. 
Such constants having the dimensions ${\rm m^3s^2kg^{-1}}$ and ${\rm   m^4s\Cdot kg^{-1}}$ are, accurate to the factor $8\pi/3$, as follows
\begin{equation}
   \varepsilon_{g} = G\Cdot\omega_0^{-4}, \hskip24pt \mu_{g} = \varepsilon_{g}c.
\label{eq=51}
\end{equation}
Here $G\approx6.674\Times10^{-11}~{\rm m^{3}kg^{-1}s^{-2}}$ is the gravitational constant and 
$\omega_0 = m{\mathit v}^2/2\hbar$ is the de Broglie circular frequency for a particle with mass $m$. For thermal neutron, for example,  $m\approx1.675\Times10^{-27}~{\rm kg}$ and ${\mathit v}\approx2\Times10^3~{\rm m/s}$ we have $\omega_0\approx5\Times10^{12}~{\rm s^{-1}}$.
From here we find
$\varepsilon_g\approx10^{-61}~{\rm m^3s^2kg^{-1}}$ and $\mu_g=\varepsilon_{g}c\approx3\Times10^{-53}~{\rm m^4s\Cdot kg^{-1}}$. As seen these 
 constants are small enough.
\item[$\bullet$]
The second quadratic form  reads
\begin{equation}
\hskip24pt
 {\mathbb F}_{\Xi\Omega}^{\,\bf T}{\mathbb F}_{\Xi\Omega} =
 {{(\Omega^2-\Xi^2)}{}}\eta_0 - 2{\bf i}({\vec \Xi}\!\cdot\!{\vec \Omega})\eta_0.
\label{eq=52}
\end{equation}
It has two invariants relative to the Lorentz transformations.
The first invariant, $I_1=(\Omega^2-\Xi^2)$, is the scalar  and the second invariant, $I_2 = 2({\vec \Xi}\!\cdot\!{\vec \Omega})$, is the pseudoscalar.
\end{itemize}

\section{\label{sec5}\Schrodinger, vorticity, and wave equations} 

In order to understand where the common force ${\vec{\mathcal F}}$ comes from, we need to write out the force density~(\ref{eq=25}) in detail. For the sake of simplicity further we will deal with the non-relativistic limit. The force density $\vec\Xi$ in this limit looks as follows
\begin{equation}
   {\vec\Xi} = {{\partial \vec p}\over{c\partial t}} + \nabla\epsilon 
   = \rho_{M} \Bigl(
                        {{\partial \vec{\mathit v}}\over{\partial t}} + \nabla{{{\mathit v}^2}\over{2}}  
                    \Bigr)
  = {\vec{F}}.                    
\label{eq=53}
\end{equation}
Here we avoid differentiation of $\rho_{M}$. Its variations are described by the continuity equation~(\ref{eq=5}).
The force term $\vec{F}$ contains all external and internal forces acting on the considered object, see the text after Eq.~(\ref{eq=33}).

The external force is gravitational because of absence of other external forces. 
From recent  full-mission Planck observations~\cite{Ade2016} it was found that  spatial curvature of the visible universe is  about 
$|\Omega_K| < 0.005$. Based on these indications we will consider a flat Euclidean space\footnote[5]{
According to Shipov~\cite{Shipov1998},  a simplest geometry with torsion, built on a variety of orientable points (points with spin) is the geometry of absolute parallelism. }.
The gravitational force we believe is represented by gradient of the gravitational potential $-\rho_{M}\nabla\phi$~\cite{SbitnevFedi2017}.

The first internal force conditioned by the pressure as a counteraction to the external force is represented by the gradient of the quantum potential 
$Q$\footnote[6]{The quantum potential is a keystone of the quantum mechanics~\cite{Madelung1926,Bohm1952a,Bohm1952b}. 
Its computations~\cite{Fiscaletti2012,Grossing2009,Nelson1966,Rapoport2005a,Rapoport2009,Rapoport2010,Sbitnev2015c} under a variety of physical circumstances shed light on the nature of the ether underlying the quantum realm.}. Note that the quantum potential is equal to the internal osmotic pressure $P$
 divided by the density distribution $\rho$~\cite{Sbitnev2016b,Sbitnev2018b}:
\begin{equation}
  Q = {{P}\over{\rho}} = {{P_2+P_1}\over{\rho}}
     =m{{D^2}\over{2}}\Biggl(
                                           {{\nabla\rho}\over{\rho}}
                                  \Biggr)^2
     - mD^2{{\nabla^2\rho}\over{\rho}}        
     = 2m{D^2}{{\nabla^2 R}\over{R}},                                                      
\label{eq=54}
\end{equation}
where $R=\sqrt{\rho}$ and the general intrinsic pressure $P$ is the sum of two pressures $P_1$ and $P_2$. One of them represents the osmotic pressure, the other represents the  kinetic energy density. The diffusion coefficient $D$ is as follows
\begin{equation}
         D = {{\hbar}\over{2m}}={{c^2}\over{2\omega}}.
\label{eq=55}
\end{equation}
Here $\hbar$ is the reduced Planck constant. Nelson in his article~\cite{Nelson1966} has described Brownian motion by the Wiener process occurring  in the ether\footnote[7]{More detail consideration of the  random generalized Brownian motion has been performed by Rapoport in~\cite{Rapoport2005,Rapoport2007}.}. In this process the diffusion coefficient $D$ is $\hbar/(2m)$.
From these considerations Nelson has derived the \Schrodinger equation.
The second part of Eq.~(\ref{eq=55}) results from the Einstein's formulas $E=mc^2$ and $E=\hbar\omega$, where $\omega$ is a wave image of the particle according to the wave-particle duality principle~\cite{Sbitnev2015c}.

It should be noted that the Brownian motion~\cite{Nelson1967,Nelson2012} takes place in the normal fluid. But we are considering the quantum processes occurring in the superfluid quantum space represented by the Bose-Einstein condensate. Observe that to deal with the ideal superfluid, we should cancel the dissipative term in the right part of  Eq,~(\ref{eq=53}). But it is not a good idea, leading to appearance of singularities. The Bose-Einstein condensate consists of two fluids~\cite{GriffinEtAl2009}. The one fluid is the superfluid. And the other fluid is the normal. These two fluids relate each other as a basin filled by the superfluid, and the background fluid playing the role of resistor. 

Therefore we remain the viscosity term in Eq,~(\ref{eq=53}). In general, this term is the viscous stress tensor, containing (mostly shear) viscosity~\cite{Romatschke2010a,Romatschke2010b,Sbitnev2015c}. Here we consider a simple dissipative term - the viscosity term $\mu\nabla^{2}{\vec{\mathit v}}$, where $\mu$ is the dynamical viscosity coefficient that fluctuates about zero in time. We believe that it is zero in the average in time, but its variance in not zero\footnote[8]{In a general case we should define the viscous stress tensor~\cite{DisconziEtAl2017,Romatschke2010a,Romatschke2010b,Sbitnev2015c}. At the transition to the nonviscous superfluid we come to the noise tensor~\cite{Rapoport2005}  fluctuating about zero.}:
\begin{equation}
 \langle \mu(t) \rangle = 0_{+}, \hskip24pt
 \langle \mu(t)\mu(0) \rangle > 0.
\label{eq=56}
\end{equation}
Here subscript ``+'' means that the viscosity coefficient is not exactly zero but has a  tiny positive value. However this value can be disclosed only on giant cosmic scales~\cite{SbitnevFedi2017}.

As a result, the force term ${\vec{F}}$ in the right part of Eq,~(\ref{eq=53}) can look as follows:
\begin{equation}
 {\vec{F}} = -\rho_{M}\nabla\phi -\rho\nabla Q + \mu(t)\nabla^2 {\vec{\mathit v}}.
\label{eq=57}
\end{equation}
Dimension of this expression is ${\rm kg}\Cdot{\rm m}^{-2}\Cdot{\rm s}^{-2}$. 
By substituting this expression in Eq.~(\ref{eq=53}) we come to the modified Navier-Stokes equation~\cite{Sbitnev2016b,Sbitnev2018b} with the shear viscosity term (the fluid layers slide on each other with friction).\\

\subsection{\label{sec51}\Schrodinger equation}

Let there is a scalar field marked by $S$.
Then the irrotational velocity ${\vec{\mathit v}}_{\rm s}$
of a particle having  the mass $m$ is defined as $\nabla S/m$. Next, the particle momentum and its kinetic energy have the following
representations:
\begin{equation}
        {\vec p} = m{\vec{\mathit v}} = \nabla S + m {\vec{\mathit v}}_{\rm o},
\label{eq=58}
\end{equation}
\begin{equation}
         m{{{\mathit v}^2}\over{2}} = {{1}\over{2m}} (\nabla S)^2 + m{{{\mathit v}_{\rm o}^2}\over{2}}
\label{eq=59}
\end{equation}
These formulas connect the irrotational field of velocities with the scalar field $S$. They permit to reduce the Navier-Stokes equation to the Hamilton-Jacobi one.
Note that the orbital velocity ${\vec{\mathit v}}_{\rm o}$ represented in these equations comes from the vorticity equation.
This equation will be considered  further.

By applying the nabla operator  to equation~(\ref{eq=53})  and taking into account the results given in Eqs.~(\ref{eq=58})-(\ref{eq=59}), we extract an equation for  the action function $S$ representing the scalar field~\cite{Sbitnev2018b}
\begin{equation}
  {{\partial}\over{\partial t}} S +{{1}\over{2m}}(\nabla S)^2 + {{m}\over{2}}{\mathit v}_{\rm o}^2
  +m\phi(\vec r) + 
  \underbrace{Q -\nu(t)\nabla^2 S}
  = C_0.
\label{eq=60}
\end{equation}
 Here $\nu(t) = \mu(t)/\rho_{M}$ is the kinematic viscosity coefficient. 
 Its dimension is the same as that of the diffusion coefficient, ${\rm m}^2\Cdot{\rm s}^{-1}$.
 Arbitrary constant $C_0$ is conditioned by the fact that this equation is the result of solving the equation 
 $(\nabla\Cdot{\vec\Xi})=4\pi\wp=(\nabla\Cdot{\vec F})$,  Eqs.~(\ref{eq=42}) and~(\ref{eq=34}).
At discarding two last terms,  enveloped by the brace, this equation is known as the Hamilton-Jacobi equation.  For pair to this equation let us also add the continuity equation
\begin{equation}
 {{\partial\rho(\vec r,t)}\over{\partial t}} +(\nabla\rho(\vec r,t)\Cdot{\vec{\mathit v}}(\vec r,t))=0.
\label{eq=61}
\end{equation}
The first equation, Eq.~(\ref{eq=60}), describes a mobility of the particle of mass $m$ in the vicinity of the point $\vec r$ at the moment of time $t$. While the second equation, Eq.~(\ref{eq=61}), describes the probability of detection of this particle in the vicinity of the point $\vec r$ at the moment of time $t$.

Since Eq.~(\ref{eq=60}) contains the quantum potential $Q$ we can go from  the two  Eqs.~(\ref{eq=60}) and~(\ref{eq=61}) to the \Schrodinger-like equation~\cite{Sbitnev2018b} 
\begin{equation}
  {\bf i}\hbar {{\partial \psi}\over{\partial t}}
  = {{1}\over{2m}}\Bigl(
                            -{\bf i}\hbar\nabla + m{\vec{\mathit v}}_{\rm o}
                            \Bigr)^2 \psi + m\phi(\vec r)\psi
  +\underbrace{m\nu(t){{d}\over{d t}}\ln(|\psi|^2) \Cdot\psi}
   - C_0\psi.                           
\label{eq=62}
\end{equation}
as soon as we define the complex wave function in the polar form
\begin{equation}
  \psi(\vec r,t) = \sqrt{\rho(\vec r,t)}\Cdot\exp\Bigl\{
                                                                          {\bf i}S(\vec r,t)/\hbar
                                                                  \Bigr\}.
\label{eq=63}
\end{equation}

  Eq.~(\ref{eq=62}) at omitting the term covered by the curly bracket represents the classical \Schrodinger equation which, even in such a simplified form, can give interesting solutions such as
\begin{equation}
 | \psi(z,y)\rangle ={{1}\over{N\sqrt{ 1 + {\bf i} \displaystyle {{\lambda_{\rm dB}z}\over{2\pi b^2}} }}}
 \sum\limits_{n=0}^{N-1}
 \exp
   \matrix{
       \left\{ -\displaystyle
         {{\Biggl(
              y - \Biggl(
                      n - \displaystyle {{N-1}\over{2}}
                   \Biggr)d
            \Biggr)^2}\over
            {2b^2\Biggl(
                         1 + {\bf i} \displaystyle {{\lambda_{\rm dB}z}\over{2\pi b^2}}
                     \Biggr)}}
       \right\}
             }.
\label{eq=64}
\end{equation}
\begin{figure}[htb!]
  \begin{picture}(200,210)(0,0)
      \includegraphics[scale=0.93]{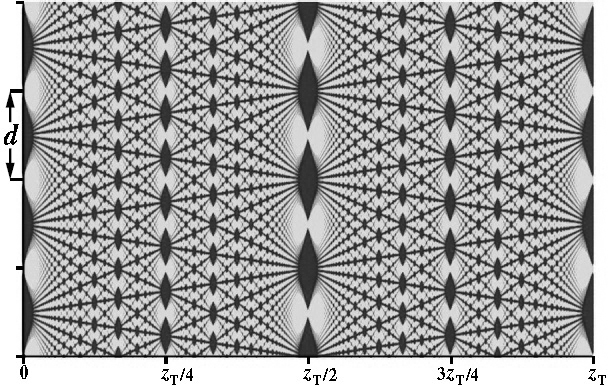}
  \end{picture}
  \caption{
 The density distribution $p(z,y) = \langle \psi(z,y)|\psi(z,y) \rangle$ shows the Talbot carpet arising when $N$ and $d$ tend to infinity.
 }
  \label{fig=7}
\end{figure}  
The solution was obtained by applying the Feynman path integral technique~\cite{Sbitnev2010b}.
It is a famous pattern of the Talbot carpet shown in  Fig.~\ref{fig=7}. It represents  a fractal, when both $N$ and $d$ approach infinity~\cite{BerryEtAl2001}.  
Here $N$ and $d$ are amount of the coherent sources of radiation (slits) and the spacing between the sources, respectively. 
The other parameters  in this function are as follows: $\lambda_{\rm dB}$ is the de Broglie wavelength of the coherent radiation and $b$ is a  cross-section of the coherent source. The parameter
\begin{equation}
   z_{\rm T} = {{d^2}\over{\lambda_{\rm dB}}}
\label{eq=65}
\end{equation}
scaling the Talbot carpet is named the Talbot length. 

\subsubsection{\label{sec511}Deceleration of a baryon-matter object in the cosmic space}

The term covered by the brace in Eq.~(\ref{eq=62}) is the Langevin's term fluctuating about zero. It represents a nonlinear color noise source. 
Since $\langle\nu(t)\rangle = 0_+$ (the average in time is slightly larger then zero), the cyclic frequency, $\omega_0 = E/\hbar$, of the particle having the energy $E=mv^2/2$ shifts to the low frequency range when the particle travels through the cosmic space:
\begin{equation}
 \psi({\vec r},t) \approx \psi_0({\vec r},t)
 \Cdot\exp\Biggl\{\!
                               -{\bf i} \omega_{0} t
                               \biggl(\!
                                        1 - {{2}\over{t}} \int\limits_0^t\!
                                        {{\nu(\tau)}\over{c^2}} {{d\ln(|\psi_0(\vec r,\tau)|^2)}\over{d\,t}}d\tau
                              \!\biggr)         
                    \!\Biggr\}.\!\!
\label{eq=66}
\end{equation} 
The particle loses the energy even in an absolute vacuum~\cite{Vigier1988} because of its scattering on the Bose-Einstein condensate that along with the superfluid strongly subjected to quantum effects~\cite{Volovik2005,Vinen2006,TsatsosEtAl2016} contains also small percent of the normal fluid~\cite{GriffinEtAl2009}.

Let the integral under the exponential be linked with the extended Hubble parameter $H_{\Lambda}$
  as follows~\cite{SbitnevFedi2017}:
\begin{equation}
\label{eq=67}
 {{2}\over{t}} \int\limits_0^t\!
  {{\nu(\tau)}\over{c^2}} {{d\ln(\rho_{m})}\over{d\,t}}d\tau \,=\,
   H_{\Lambda}t. 
\end{equation}
 The extended Hubble constant, $H_{\Lambda}$, reads
\begin{equation} 
   H_{\Lambda}  = \sqrt{H_{0}^{2}\;+\;\Lambda{{c^2}\over{3}}}, ~~~
   H_{0} = \sqrt{ {{8\pi G\rho_{m}}\over{3}} } . 
\label{eq=68}
\end{equation}
Here we are dealing with a flat universe~\cite{Ade2016}, therefore only two terms are represented under the root:  the original Hubble parameter, $H_{0}\approx2.36\times10^{-18}~{\rm s^{-1}}$, and the correcting term,  $\Lambda c^2/3$, as resulting from the first Friedman equation~\cite{Friedman1922}, where $c$ is the speed of light and the cosmological constant $\Lambda$ is about $10^{-52}~{\rm m^{-2}}$~\cite{DasBhaduri2015}. So, from Eq.~(\ref{eq=68}) we obtain $H_{\Lambda}\approx 2.93\times10^{-18}~{\rm s^{-1}}$.

In the eikonal approximation~\cite{Rapoport2009} the acceleration of a baryon-matter object, traveling through the superfluid quantum space, can be written in accordance with Newton's second law
\begin{equation}
 {\vec a} = {{d{\vec{\mathit v}}}\over{dt}} = \nu(t)\nabla^2{\vec{\mathit v}}
 = -\nu(t)\nabla\Biggl(
                                {{d\ln(\rho_{m})}\over{dt}}
                       \Biggr).
\label{eq=69}
\end{equation}
Here  $d\ln(\rho_{m})/dt$ is equal to $-(\nabla\cdot{\vec{\mathit v}})$, what follows from the continuity equation~(\ref{eq=61}) for the mass density 
distribution $\rho_m({\vec r},t) =m\rho({\vec r},t)$, namely
\begin{equation}
\hskip-8pt
  {{d\ln(\rho_{m})}\over{dt}} \!=\!
  {{\partial\ln(\rho_{m})}\over{\partial t}} + ({\vec{\mathit v}}\cdot\!\nabla\!\ln(\rho_{m})) \!=\! -(\nabla\!\cdot\!{\vec{\mathit v}}).
\label{eq=70}
\end{equation}

 The right part in Eq.~(\ref{eq=69}) can be replaced by the combination of the Hubble parameters $H_0$ and $H_{\Lambda}$, as soon as we multiply Eq.~(\ref{eq=67}) by $tc^2/2$ and differentiate it with respect to $t$. We get
\begin{equation}
 {\vec a} = -\nabla tc^2 H_{\Lambda} - \nabla {{t^2c^2}\over{4H_{\Lambda}}}H_{0}^{2}{{d\ln(\rho_{m})}\over{dt}}
\label{eq=71}
\end{equation} 
where the operator $\nabla = d/d\ell$ comes from Eq.~(\ref{eq=69}) and calculates a gradient along the path $\ell$. Let $d\ell/dt$ be the speed of light. Then we get
\begin{equation}
  a = - H_{\Lambda}c
      \Biggl(\!
              1 +  {{H_0^2}\over{H_{\Lambda}^2}}
               \biggl(
                        {{t}\over{2\rho_{m}}} {{d\rho_{m}}\over{dt}}
              \biggr)
      \!\Biggr)
\label{eq=72}
\end{equation} \\
 The continuity equation reads $d\rho_{m}/dt=0$, see Eq.~(\ref{eq=5}). 
 From there it follows that the term under the inner brackets vanishes. We finally obtain
\begin{equation}
   a = - H_{\Lambda}c \approx -8.785\times10^{-10}~{\rm m\cdot s^{-2}}. 
\label{eq=73}
\end{equation}

 This  indicates a good agreement with the acceleration $a_P=(8.74\pm1.33)\times10^{-10}~{\rm m\cdot s^{-2}}$, which became known as the Pioneer anomaly~\cite{AndersonEtAl1998}. 
 And what is no less important, the computations show a negative sign at the number, what means that the acceleration pushes the spacecraft back towards the Sun.  

In that regard of particular interest is the paper of Sarkar, Vaz, and Wijewardhana~\cite{SarkarEtAl2018} devoted to development of a self-consistent, Gravitoelectromagnetic formulation of a slowly rotating, self-gravitating and dilute Bose-Einstein condensate (BEC), intended for astrophysical applications in the context of dark matter halos. 
It is noteworthy, that the authors come to the nonlinear \Schrodinger equation and further by employing the Madelung transformation come to equations describing behavior of the ideal fluid that is the superfluid BEC. 
Authors propose  as an alternative to cold dark matter a light boson whose mass is small enough (as low as 10$^{-33}$ eV). Its critical temperature is well above that of the Cosmic Microwave Background. This can ensure that a significant fraction of the bosons settle in the ground state with forming the superfluid BEC.

\subsection{\label{sec52}Vorticity equation}

The vorticity equation results from the second gravitomagnetic equation, from Eq.~(\ref{eq=41}):
\begin{equation}
 {{1}\over{c}}{{\partial}\over{\partial t}}{\vec\Omega} =  [\nabla\Times{\vec{F}}]
 \hskip14pt
 \Rightarrow
 \hskip14pt
 \rho_{M}{{\partial\vec\omega}\over{\partial t}} = \mu(t)\nabla^2\vec\omega
\label{eq=74}
\end{equation}
Here we have replaced $\vec\Xi$ by $\vec{F}$, as shown in Eq.~(\ref{eq=53}).
As for the second equation, we do not differentiate the mass density $\rho_{M}$ by time in the expression $\vec\Omega=c\rho_{M}\vec\omega$. 
Also we have $[\nabla\Times{\vec{F}}]=\mu(t)[\nabla\Times\nabla^2{\vec{\mathit v}}]= \mu(t)\nabla^2\vec\omega$.
The vorticity equation describes vortex motion in a local reference frame sliding along an optimal trajectory guided by the wave function that is solutions of the \Schrodinger equation~(\ref{eq=62}). In this case the vortex ideally simulates, in the eiconal approximation, the particle moving along the Bohmian trajectory~\cite{Sbitnev2018b}. 

\begin{figure}[htb!]
  \begin{picture}(200,110)(0,0)
      \includegraphics[scale=0.5]{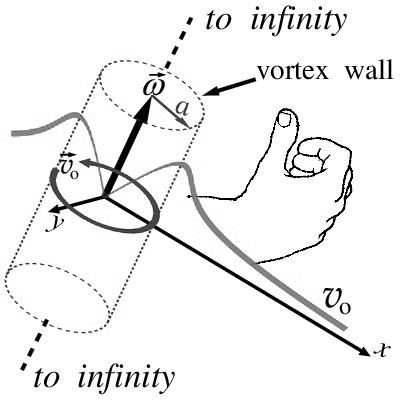}
  \end{picture}
  \caption{
  Vortex given in the cylindrical coordinate system. 
  The vortex wall is a boundary, where the orbital speed, ${\mathit v}_{o}$, reaches maximal values;
  $a$~is the radius of the vortex tube.
 }
  \label{fig=8}
\end{figure}

For the purpose to study the vortex dynamics in 3D space, spherical and toroidal coordinate systems are ideally suited. In order to show some specific features of vortices in their cross-section, the cylindrical coordinate system is better employed, Fig.~\ref{fig=8}. 
In this figure the vortex tube is oriented along the $z$-axis and its central line is placed in the  origin, $(x=0, y=0)$. Let us look on the vortex tube in its cross-section.  Equation~(\ref{eq=74}), written down for the cross-section of the vortex, looks as follows
\begin{equation}
 {{\partial \omega}\over{\partial t}} = 
  \nu(t)\Biggl(
            {{\partial^2\omega}\over{\partial r^2}} 
             +{{1}\over{r}}{{\partial \omega}\over{\partial r}}
          \Biggr).
\label{eq=75}
\end{equation}
 Here $\nu(t) = \mu(t)/\rho_{M}$ and $\omega$ means value of the vorticity lying along $z$-axis.

 A general solution of this equation has the following view:
\begin{eqnarray}
\label{eq=76}
\hskip-16pt
\omega(r,t) &=& {{\mit\Gamma}\over{4\mit\Sigma(\nu,t,\sigma)}}
\exp\Biggl\{
         -{{r^2}\over{4\mit\Sigma(\nu,t,\sigma)}}
      \Biggr\}, 
\\ 
\hskip-16pt
  {\mathit v}(r,t) &=& {{1}\over{r}}\int\limits_0^r \omega(r',t)r'dr'
  \;=\; {{\mit\Gamma}\over{2r}}
  \Biggl(
     1 - \exp\Biggl\{
             -{{r^2}\over{4\mit\Sigma(\nu,t,\sigma)}}
                \Biggr\}
  \Biggr).
\label{eq=77}  
\end{eqnarray}
 Here $\mit\Gamma$ is the integration constant having dimension [length$^2$/time] 
 and the denominator $\mit\Sigma(\nu,t,\sigma)$ has a view
\begin{equation}
 \mit\Sigma(\nu,t,\sigma) =
 \int\limits_0^t \nu(\tau)d\tau + \sigma^2.
\label{eq=78}
\end{equation} 
 Here $\sigma$ is an arbitrary constant such that the denominator is always positive.
 
 The orbital velocity, ${\vec{\mathit v}}_{o}$ shown in Fig.~\ref{fig=8}, is drawn by thick gray curve  in the plane $(x, z)$. A maximal value of this curve crosses the boundary named by the vortex wall. The vortex wall outlines a region named by the vortex core, where the orbital velocity falls out to zero at moving to the center of the vortex. Also it falls to zero at removing from the wall to infinity.

 Fig.~\ref{fig=2}(a) shows a torus formed by gluing opposite ends of the cylindrical vortex. Observe that the torus surface  is the vortex wall where the orbital speeds are maximal. 
Position of points on the toroidal vortex wall with the tube radius $a$ and the torus radius $b$  is given  in the Cartesian coordinate system by the following set of equations~\cite{Sbitnev2016c}:
\begin{equation}
\hskip-16pt
  \left\{
     \matrix{
         x = (b + a\cos(\varpi_0t+\phi_0))\cos(\varpi_1t+\phi_1), \cr
         y = (b + a\cos(\varpi_0t+\phi_0))\sin(\varpi_1t+\phi_1), \cr
         z = a\cos(\varpi_0t+\phi_0).\hskip86pt \cr
               }
  \right.
\label{eq=79}
\end{equation}
 Here the frequency $\varpi_0$  is that of rotation about  the center of the tube pointed by letter $c$ in Fig~\ref{fig=2}(a).
 The frequency $\varpi_1$ is that of rotation about the center of the torus localized in the origin of  coordinates.
 The  toroidal vortex wall can be filled by helicoidal strings everywhere densely at choosing different phases $\phi_0$ and $\phi_1$ ranging from $0$ to $2\pi$ with the infinitesimal  increment~\cite{Sbitnev2016c}.
One string is shown in this figure marked by symbol ${\mathit S}_1$. It is drawn when we choose   $\varpi_1 =  \varpi_0/2$ and $\phi_0 =\phi_1 = 0$.
Arrows along the strings show directions of a current, flowing  when $t$ goes up.

\subsubsection{\label{sec521}Neutron spin-echo spectroscopy:  measuring the gravitomagnetic effect}

Some years ago, a fine experiment was conducted with ultra-cold neutrons flying over a flat mirror elastically reflecting the neutrons in the gravitational field of the Earth~\cite{NesvizhevskyEtAl2002}. Recently other experiment was conducted with ultra-cold neutrons by applying the resonance spectroscopy to gravity~\cite{JenkeEtAl2011,JenkeEtAl2014}. An important detail of the experiments were absence of the electromagnetic field, whose force is several orders of magnitude greater than the gravitational field. For this reason, it was possible to observe the quantum effects of flying neutrons in the gravitational field.

Observation of gravitomagnetic effects at presence of the electromagnetic field opens a possibility to study spin and torsion effects in a gravitational field~\cite{RapoportSternberg1984a,RapoportSternberg1984b,Sternberg1985} with implications for particle physics, cosmology etc.
It requires using of sensitive experimental devices. Among them the spin-echo spectroscopy is taken a special place, since the spin echo spectrometers possess by an extremely high energy resolution. 

\begin{figure}[htb!]
  \begin{picture}(170,280)(0,0)
      \includegraphics[scale=0.6,angle=0]{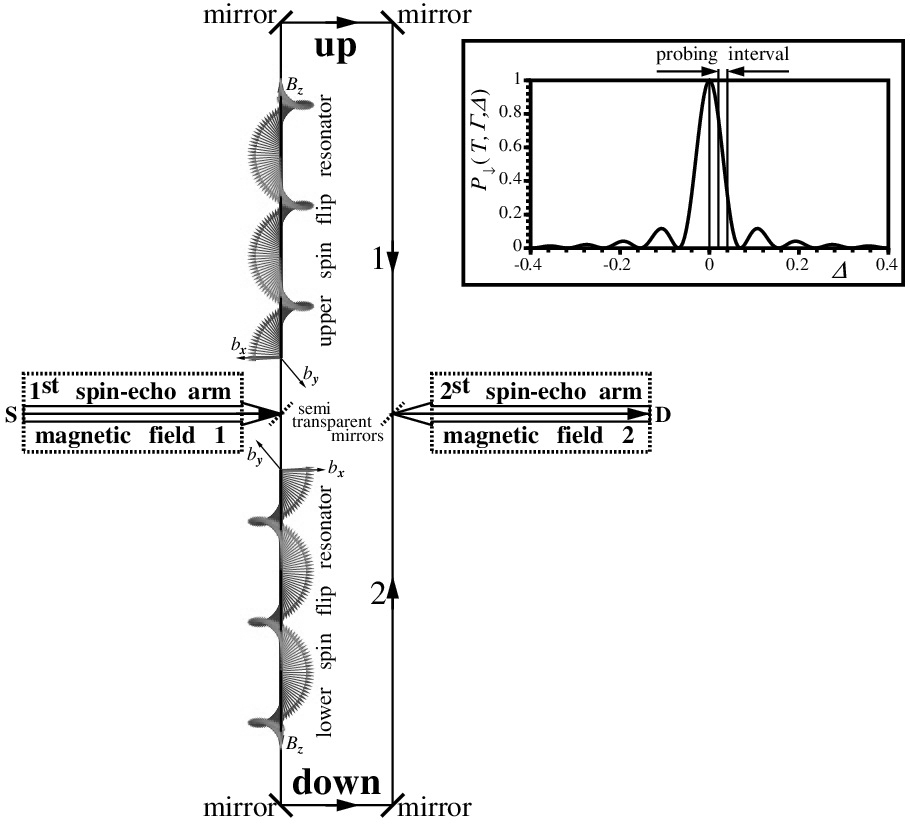}
  \end{picture}
  \caption{
The spin-echo  spectrometer (placed in horizontal direction) contains the spin resonators with opposite oriented periodic magnetic structures as sample (placed vertically). 
Thin arrows  show paths 1 and 2 of the polarized neutrons. Thick empty arrows show orientations of magnetic fields 1 and 2 in the arms of the spin-echo spectrometer. S and D are the neutron source and the detector.
Insert in the upper right shows the resonance curve~(\ref{eq=83}). Here the interval for probing falls on a steep area of the spin resonance.
              }
  \label{fig=9}
\end{figure}

A spin-echo spectrometer for the first time proposed by Ferenc Mezei~\cite{MezeiEtAl2003} in the 1970s is a very useful realization demonstrating Bohm's idea of the wholeness and implicate order~\cite{Bohm2006}. In the spin-echo spectroscopy experiment a polarized neutron beam unfolds on the first spin-echo arm and on the second arm it folds, see Fig.~\ref{fig=9}.  If we remove a sample (here it is represented by two vertical arms of the spatial spin flip resonator) then after the second spin-echo arm it is observed that the initial polarization of the neutron beam is restored.

Analogous unfolding and folding phenomenon was described by David Bohm at observing rotation forwards and backwards of a large rotating cylinder specially embedded in a jar~\cite{Bohm2006}. Here is what  Michael Talbot writes in his book "The Holographic Universe"~\cite{Talbot1991}:
"The narrow space  between the cylinder and the jar  was filled with glycerine - a thick, clear liquid - and floating motionlessly in the glycerine  was  a  drop  of  ink.  What  interested   Bohm  was  that   when  the handle on the cylinder was turned, the drop of ink spread  out  through the  syrupy  glycerine  and  seemed  to  disappear.  But  as  soon  as  the handle was  turned  back  in the  opposite direction,  the  faint tracing of ink  slowly collapsed  upon itself  and once again  formed  a  droplet". 

The presence of the sample in the spin-echo spectrometer introduces a disturbance in the polarization pattern of neutrons, the result of which is amplified greatly after passing through the second arm.

We propose here  an experiment with polarized neutrons passing through the spin magnetic resonators oriented in opposite directions. One resonator is oriented up and and other down, Fig.~\ref{fig=9}. These resonators are adopted as a sample placed between two arms of the spin-echo spectrometer.
After passing the resonators the both beams on the mirrors unfold back and then they are directed through the second spin-echo arm.  
The aim is to observe difference of the topological phases arising at the spin resonance due to different  neutron speeds induced by the gravitational acceleration of Earth.

Before going further, let us make a number of estimates for the case of thermal neutrons, $E\approx 0.025~{\rm eV}$, passing through magnetic periodic structures, each with a path length $L \sim 1~{\rm m}$. The neutron speed ${\mathit v}$ is about $2\Cdot10^{3}~{\rm m/s}$ and the time of flight, $\tau$, is about $5\Cdot10^{-4}~{\rm s}$. 
At vertical flight of a thermal neutron in the gravitational field of Earth its speed varies slightly  by a value of $g\tau \approx 4.9\Cdot10^{-3}~{\rm m/s}$, where $g\approx9.807~{\rm m\Cdot s^{-2}}$ is the gravitational acceleration of Earth. 
At flight of the neutron to up and down we find changes of the speed:
\begin{equation}
 {\mathit v}_{1,2} = {\mathit v} \mp g\Cdot\tau \approx
 \cases{1.9999951 \cr 
            2.0000049 \cr
           } {\Times10^3~~\rm{{m}\over{s}}}.
\label{eq=80}
\end{equation}
Variations of the speed $\mathit v$ are seen to be observed in 7th sign after dot. In order to register such variations we need to use the spin resonance technique with applying the spin-echo effect.

For simplicity we will consider a helical magnetic configuration~\cite{IoffeEtAl1991,Sbitnev2018c}:
\begin{equation}
{\vec B} =
\{
  b\cos(\omega t), b\sin(\omega t), B_z =~{\rm const}
\}.
\label{eq=81}
\end{equation}
Here $B_z$ is the principal magnetic field and $b$ is a small transversal magnetic field, $b\ll B_z$, and $\omega$ is the frequency of oscillations of the transversal component. In our evaluations $\omega = 2\pi{\mathit v}/L \approx 1.256637\Cdot10^{4}~{\rm s^{-1}}$ and
\begin{equation}
 \omega_{1,2} = {2\pi}\Biggl(
                                          {{\mathit v}\over{L}} \mp {{g}\over{\mathit v}}
                                  \Biggr)  \approx
 \cases{1.256634 \cr
           1.256640 \cr
           } {\Times10^4~\rm{s^{-1}}}.
\label{eq=82}
\end{equation}

The spin-flip resonance curve (turn spin down probability) for neutrons 
passing through the both resonators has a standard form~\cite{AgamalyanEtAl1988,Sbitnev2018c}:
\begin{equation}
   P_{\downarrow}(\tau,{\mit\Gamma},{\mit\Delta_k})
 = |\phi_{\downarrow}|^2 =   s_x^2(\tau) + s_y^2(\tau) 
 =   {{\mit\Gamma^2}\over{\mit\Gamma^2+\mit\Delta_k^2}}
   \sin^2\biggl(
      {{\tau}\over{2}}\sqrt{\displaystyle{\mit\Gamma^2+\mit\Delta_k^2}}
            \biggr).
\label{eq=83}
\end{equation} 
 Here $k$ ranges 1, 2 and the parameters
\begin{equation}
  {\mit\Delta_k}     = \omega_k - {\mit\Omega}_*\cos(\theta),~~~~
  {\mit\Gamma} = {\mit\Omega}_*\sin(\theta)
\label{eq=84}
\end{equation} 
 specify the position and the width of a resonance maximum, where
\begin{equation}
{\mit\Omega}_* = \gamma_n \sqrt{B_z^2 + b^2},~~~~
\theta = \arctan(b/B_z)
\label{eq=85}
\end{equation} 
 are the Larmor precession frequency and the apex angle of the cone described by the vector $\vec B$, respectively~\cite{IoffeEtAl1991}.
 
One can see that the detuning of the resonances for $k =1, 2$  has the following value
\begin{equation}
 \delta\omega\Cdot\tau= ({\mit\Delta}_2 - {\mit\Delta}_1)\tau = 
 4\pi{{g}\over{\mathit v}}\Cdot\tau \approx3\Cdot10^{-5}.
\label{eq=86}
\end{equation}
 From here it follows that than smaller the neutron speed $\mathit v$, the larger $\delta\omega$.
That is why, for such experiments it is better to take cold neutrons and even ultra-cold ones.
On the other hand, one can  try to make an acute resonance curve by choosing the inequality $B_z \gg b$ as large as possible. 
Observation of such small effects in this case  is done on a a steep slope of the resonance curve, see insert in Fig.~\ref{fig=9}.
Attracting the spin-echo spectrometry for detecting such small phase variations is an extra effective method for the experimental testing.

\subsubsection{\label{sec522}Spiral galaxy rotation}
  
 Since the time average of the viscosity is zero, according to our agreement~(\ref{eq=56}), the above solutions, Eqs.~(\ref{eq=76}) and~(\ref{eq=77}), at large times, $t\gg1$ reduce to:
\begin{eqnarray}
\label{eq=87}
 \omega(r,\sigma) &=& {{\mit\Gamma}\over{4\sigma^2}}
                         \exp\Biggl\{
                                  -{{r^2}\over{4\sigma^2}}
                               \Biggr\},
\\                               
 {\mathit v}(r,\sigma) &=& {{\mit\Gamma}\over{2r}}
                 \Biggl(
                  1 -  \exp\Biggl\{
                                  -{{r^2}\over{4\sigma^2}}
                               \Biggr\}
                   \Biggr) .           
\label{eq=88}
\end{eqnarray} 
 This solution is named the Gaussian coherent vortex cloud~\cite{KleinebergFriedrich2013,NegrettiBillant2013,ProvenzaleEtAl2008}  that is permanent in time. 
\begin{figure}[htb!]
  \begin{picture}(120,70)(0,0)
      \includegraphics[scale=0.45]{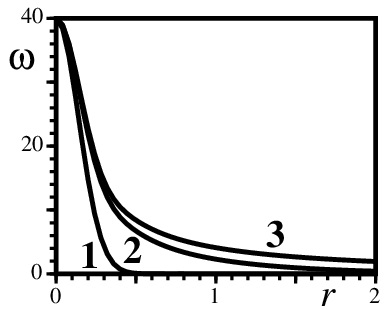}
  \end{picture}
  \caption{
 Vorticity~(\ref{eq=89}) as a function of $r$ for different sums of $\sigma\Cdot n$ ($\sigma=0.1$): (1) $N=1$, (2) $N=10$, (3) $N=100$. The vorticity is calibrated by height at the point $r=0$  by choosing $\mit\Gamma$. (The scales are conditional).
 }
  \label{fig=10}
\end{figure}  
\begin{figure}[htb!]
  \begin{picture}(120,150)(0,0)
      \includegraphics[scale=0.40]{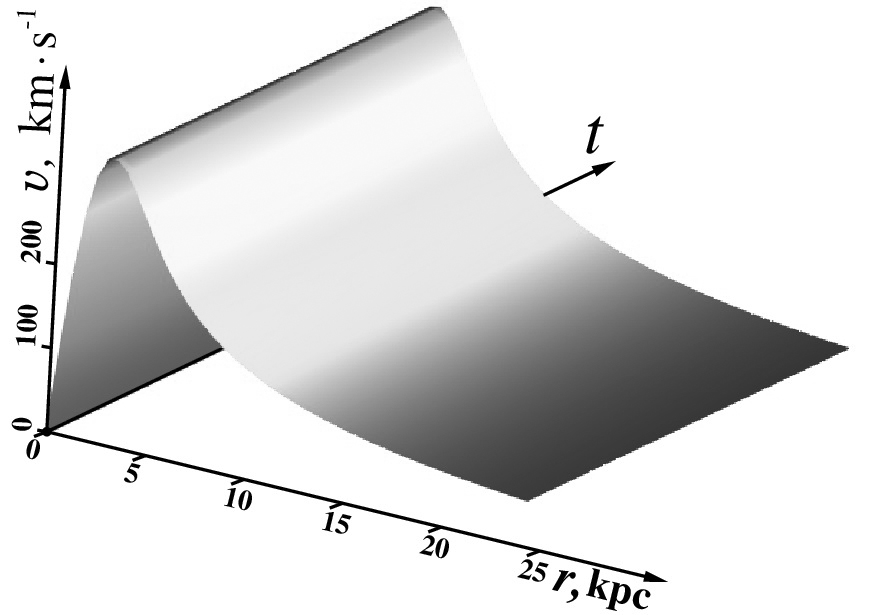}
  \end{picture}
  \caption{
 Orbital velocity~(\ref{eq=90}) for $N = 1$
 }
  \label{fig=11}
  \begin{picture}(120,165)(0,0)
      \includegraphics[scale=0.40]{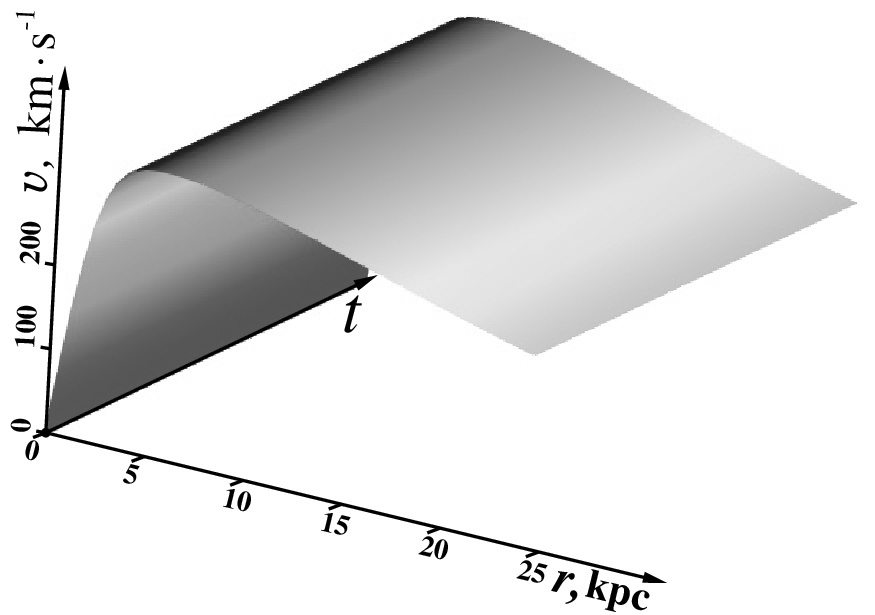}
  \end{picture}
  \caption{
 Orbital velocity~(\ref{eq=90}) for $N = 10$
 }
  \label{fig=12}
  \begin{picture}(120,165)(0,0)
      \includegraphics[scale=0.40]{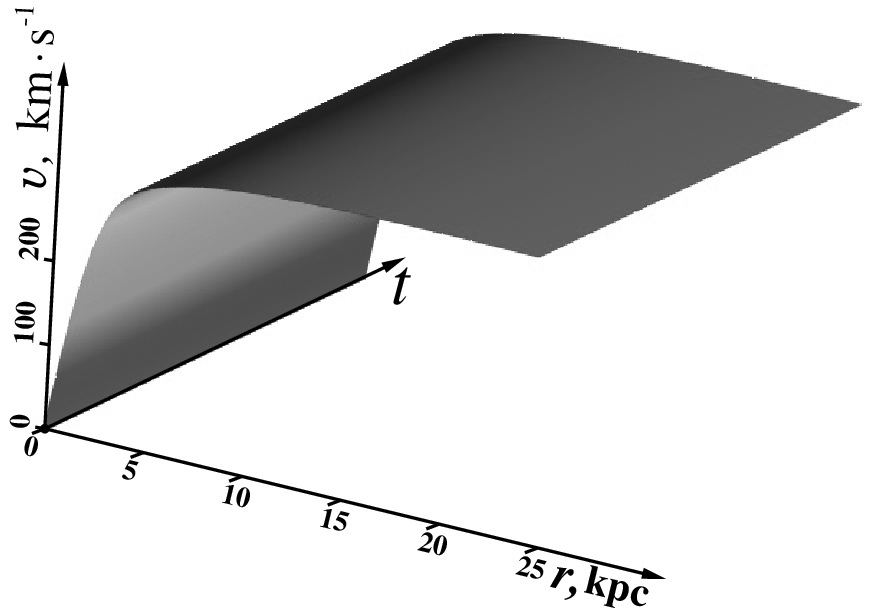}
  \end{picture}
  \caption{
 Orbital velocity~(\ref{eq=90}) for $N = 100$
 }
  \label{fig=13}
\end{figure}  

The Gaussian vortex clouds unleashed from the time dependence allow superposition
\begin{eqnarray}
\label{eq=89}
       \omega(r)  &=& \sum\limits_{n=1}^{N} \omega(r,\sigma_n), 
\\
   {\mathit v}(r) &=& \sum\limits_{n=1}^{N} {\mathit v}(r,\sigma_n),
\label{eq=90}
\end{eqnarray}
 where $\sigma_{n}$ grows with increasing $n$.
 A weak dependence on $t$ at the noise level is acceptable.
  Sums of the Gaussian vortex clouds at different set of $\sigma\Cdot n = 0.1n$, $n=1,2, \cdots, N$ are shown in Fig.~\ref{fig=10}. At $N=1$ we have a single Gaussian vortex cloud. One can see that its tail vanishes very quickly. On the other hand, at $N=100$ the tail is kept very long, slowly decreasing to zero. It means that the orbital velocity~(\ref{eq=90}) can maintain a constant level at long distances from the vortex core.
 
Figs.~\ref{fig=11},~\ref{fig=12},~\ref{fig=13} show the orbital speeds~(\ref{eq=90}) as functions of $r$ and $t$ for three different values of $N$: $N=1$, $N=10$, and $N=100$, respectively. First, all speeds are constant with regards of increasing time $t$.  As for increasing $r$, the orbital speeds show different behavior at large $r$. The orbital speed with a single $\sigma$-mode, $N=1$, has usual profile as for a standard classical vortex. Inside the vortex core the speed falls down to zero when $r$ tends to zero. There is a vortex wall where the orbital speed reaches maximal values. Further, with increasing $r$ the speed begins to quickly fall, Fig.~\ref{fig=11}. It decreases up to zero at $r$ tending to infinity. When the next $\sigma$-modes are added to the sum, $\sigma = 0.1\cdot n$, $n=1,2,\cdots, N$~(\ref{eq=90}), a profile of the orbital speed behind the vortex wall stops falling, Fig.~\ref{fig=12}. In case of $N=100$ modes, for example, the orbital speed takes a flat profile, Fig.~\ref{fig=13}.

Spiral galaxies pose different flat profiles of the orbital speeds~\cite{deBlokEtAl2001,Rubin2004} depending on $\sigma$-modes of the Gaussian vortex clouds composed from different dark matter constituents.

\subsection{\label{sec53}Wave equations}

The wave equations for the vector fields $\vec\Xi$ and $\vec\Omega$ traveling in the cosmic space stems from Eq.~(\ref{eq=39}) as soon as we apply to this equation the differential operator $\mathcal D$:
\begin{equation}
 {\mathcal D}{\mathcal D}^{\rm T} {\mathbb F}_{\Xi\Omega} =  \square {\mathbb F}_{\Xi\Omega} = {{4\pi}\over{c}}{\mathcal D}{\vec{\bJ}}
\label{eq=91} 
\end{equation}
Observe that the term ${\mathcal D}{\vec{\bJ}}$ performs the same computations that ${\mathcal D}T$ in~(\ref{eq=23}). Therefore we can rewrite the above formula in more detail
\begin{equation}
  \biggl(
      \nabla^2 - {{\partial^2}\over{c^2\partial t^2}}
  \biggr)
  ({\vec\Omega} - {\bf i}\,{\vec\Xi}) = {{4\pi}\over{c}}
  \biggl(
           [\nabla\Times{\vec\Im}]
          -{\bf i} \Bigl(
                             \nabla{c\wp}+{{\partial}\over{c\partial t}}{\vec\Im}
                     \Bigr)
  \biggr)
\label{eq=92}
\end{equation}

Now let us to take a look on the wave propagation dynamics through a prism of Eqs.~(\ref{eq=40})--(\ref{eq=43}). The energy flux~(\ref{eq=50}) can be transferred by waves described by the wave equations. They can be extracted from Eq.~(\ref{eq=43}) after applying to it either the operator curl
\begin{eqnarray}
\nonumber
 &&{{1}\over{c^2}}{{\partial^2}\over{\partial t^2}}{\vec\Omega} 
 + \overbrace{
    \nabla\underbrace{(\nabla\Cdot{\vec\Omega})
                              }_{=~0}
 - \nabla^2{\vec\Omega}           
                  }^{[\nabla\Times[\nabla\Times{\vec\Omega}]]}                   
 = -{{4\pi}\over{c}}[\nabla\Times{\vec{\Im}}]     
\\
&\Rightarrow\;\;&
 {{1}\over{c^2}}{{\partial^2}\over{\partial t^2}}{\vec\Omega} - \nabla^2{\vec\Omega}
  = -{{4\pi}\over{c}}[\nabla\Times{\vec{\Im}}] ,
\label{eq=93}
\end{eqnarray}
or by differentiating by time 
\begin{eqnarray}
\nonumber
&&
 {{1}\over{c^2}}{{\partial^2}\over{\partial t^2}}{\vec\Xi} 
 +          \overbrace{
  \nabla\underbrace{(\nabla\Cdot{\vec\Xi})
                             }_{=~4\pi\wp} 
 - \nabla^2{\vec\Xi}}^{[\nabla\Times[\nabla\Times{\vec\Xi}]]}
=   -{{4\pi}\over{c^2}}{{\partial }\over{\partial t}}   {\vec{\Im}}
\hskip24pt                          
\\
&\Rightarrow\;\;&
 {{1}\over{c^2}}{{\partial^2}\over{\partial t^2}}{\vec\Xi} - \nabla^2{\vec\Xi}
= -{4\pi}\Bigl(
                     \nabla\wp + {{\partial}\over{c^2\partial t}}{\vec\Im}
             \Bigr).
\label{eq=94}
\end{eqnarray}
First one can note, that in a free space these waves propagate freely like radio waves. In this case, the right parts in these equations should be absent.
 
However, all space is filled by the superfluid quantum medium having  the zero-point fluctuations. Or, as suggested, the space is filled by the dark energy and the dark matter. Their fluctuations at the nonzero temperature of cosmic space, equal about to $2.725~{\rm K}$, are believed to  manifest themselves  as the cosmic microwave  background (CMB), likewise to CMB of the EM radiations~\cite{LiWu2016,NozariaEtAl2018}.
 These sources are placed in the right parts of the wave equations~(\ref{eq=93})-(\ref{eq=94}).
These parts, namely,  $\wp$ and ${\vec\Im}$, reveal an amazing accordance to the electrodynamic potentials  $\phi$~and~$\vec A$.
We see, that the term
\begin{equation}
 {\vec{\mathcal B}}=(4\pi/c)[\nabla\Times{\vec\Im}]
\label{eq=95}
\end{equation}
in the right part of Eq.~(\ref{eq=93}) represents a source of the angular moment waves. And the term
\begin{equation} 
 {\vec{\mathcal E}}=(4\pi)\nabla\wp + (4\pi/c){{\partial~}\over{c\partial t}}{\vec\Im}
\label{eq=96}
\end{equation}
in the right part of Eq.~(\ref{eq=94}) represents a source of the force density waves.
 Dimension of these sources is pressure per unit volume. 

\section{\label{sec6}CMB radiation} 
   
 Note that $\wp$ is represented as divergence~(\ref{eq=34})  of all force densities~(\ref{eq=57})  acting in the unit volume under consideration:
\begin{equation}
 \wp \sim (\nabla\Cdot{\vec{F}}) = 
 \lim_{V\rightarrow 0} {{1}\over{V}}
 \oint\limits_{S} ({\vec{F}}\Cdot d{\vec S}).
\label{eq=97}
\end{equation} 
Here the integral over a closed surface $S$ bounding the volume $V$ computes a flow of the vector field ${\vec{F}}$ through this surface. The density current $\vec\Im$ is simply equal to ${\vec{\mathit v}}\wp$, where the velocity ${\vec{\mathit v}}$ is the sum of  irrotational and orbital velocities~(\ref{eq=32}).
From here one can see, that the oscillating sources push down a quantum superfluid substance from the volume $V$ through the surface $S$ in one half-period and retract it back (negative pressure) in other half-period. 
For that reason Eqs.~(\ref{eq=93}) and~(\ref{eq=94}) have the following view
\begin{eqnarray}
\label{eq=98}
  \square\; {\vec\Omega} &=& {\vec{\mathcal B}}, \\
  \square\; {\vec\Xi}        &=&  {\vec{\mathcal E}},
\label{eq=99}
\end{eqnarray}
with the right parts different from zero.
These equations describe oscillations of  vector fields, quanta of which can be gravitational  Bose particles with 
 a tiny mass~\cite{BerezhianiKhoury2015,DasBhaduri2015,MielczarektAl2010,SarkarEtAl2018,Sekine2016}. 

Fluctuations of the sources $\vec{\mathcal B}$ and $\vec{\mathcal E}$ can take place, for example, by rotating of two massive bodies about each other. 
However, weak fluctuations of the force fields $\vec{\mathcal B}$ and $\vec{\mathcal E}$ can occur continuously because of nonzero temperature of outer space equal about $2.725~{\rm K}$,  supporting the CMB radiation~\cite{LiWu2016,NozariaEtAl2018}.

Main parameters involved in Eqs.~(\ref{eq=17})-(\ref{eq=18}) come from the mass density: 
\begin{equation}
 \rho_{M} = {{M}\over{\Delta V}} = m{{N}\over{\Delta V}} = m\rho.
\label{eq=100}
\end{equation}
 The mass density is a ratio of the bulk $M$ to the volume $\Delta V$ occupied by this mass. In turn out, that there are $N$ elementary carriers of mass $m$,  which give the bulk $M$~\cite{JackiwEtAl2004}. Ratio of $N$ to the fluid volume $\Delta V$ is the density $\rho = N/{\Delta V}$. 

Let us evaluate the mass density of our visible universe. Here we will adhere units SI.  At the diameter of the universe $d$ equal about 28.5~gigaparsecs~\cite{BarsTerning2010}, or $8.8\Times10^{26}~{\rm m}$, its volume is $\Delta V = (\pi/6)d^{\,3} \approx3.57\Times10^{80}~{\rm m^3}$.
The total mass of ordinary matter in the universe according to different sources is about $10^{53}$ to $2\Times10^{53}$~kg~\cite{Davies2006,Vlahovic2011}. We adopt the average $M=1.5\Times10^{53}$~kg. 
From here we may evaluate the mass density of the ordinary matter filling the visible universe
\begin{equation}
   \rho_{M} = {{M}\over{\Delta V}}  
   \approx 
   5.6\Times10^{-28}~{\rm kg\Cdot m^{-3}}.
\label{eq=101}
\end{equation}

From the other hand, because space is nearly flat (Euclidean) today, the average density is close to the critical density~\cite{Ryden2006}:
\begin{equation}
 \rho_{\rm crit} = {{3}\over{8\pi G}}H_0^2 
 \approx 
 10^{-26}~{\rm kg\Cdot m^{-3}}.
\label{eq=102}
\end{equation}
Here $G$ is the gravitational constant. 
The Hubble constant $H_0$ expressed in units adopted in astrophysics is about 
$73~{\rm km\Cdot Mpc^{-1}\Cdot s^{-1}}$\cite{FreedmanEtAl2001}, 
or $2.36\Times10^{-18}~{\rm s^{-1}}$ being represented in SI unit~\cite{SbitnevFedi2017}.

Now we can estimate the presence of ordinary matter and the matter composing the critical density (dark matter and energy) in the observed universe in the percentage
\begin{equation}
  {{\rho_{\rm crit}}\over{\rho_{M}+\rho_{\rm crit}}}\Cdot100\% \approx 96\%
\label{eq=103}
\end{equation}
So, the ordinary (baryon) matter is only $4\%$ in the visible universe.
\begin{figure}[htb!]
  \begin{picture}(120,180)(0,0)
      \includegraphics[scale=0.45]{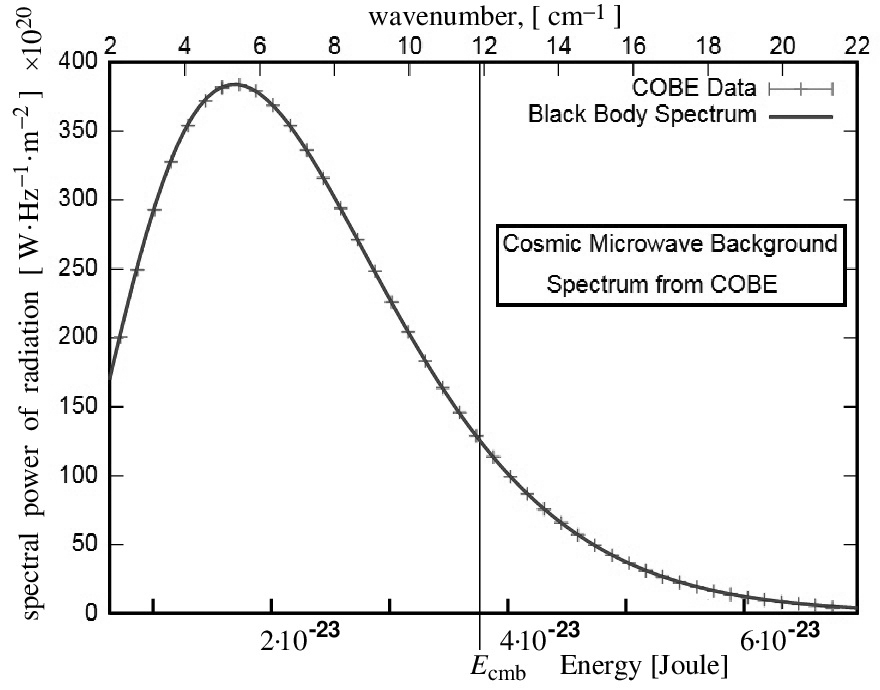}
  \end{picture}
  \caption{
 Cosmic microwave background spectrum measured by the FIRAS instrument on the COBE~\cite{CMB2018}. 
 }
  \label{fig=14}
\end{figure} 

The cosmic microwave background (CMB) radiation of the outer space is an emission of uniform, black body thermal energy coming from all parts of the sky. It has a temperature $T_{cmb} = 2.725~{\rm K}$. 
The radiation submits to the Planck's law:
\begin{equation}
  R_{\rm cmb}(\nu,T_{\rm cmb}) = {{2h\nu^3}\over{c^2}}
  {{1}\over{\displaystyle\exp\Biggr\{{{h\nu}\over{k_BT_{\rm cmb}}}\Biggl\}-1}}.
\label{eq=104}
\end{equation}
 Here $h$ is the Planck constant, $\nu$ is the frequency of the radiation, and $k_B$ is the Boltzmann constant.
 The spectral power of the radiation, $R(\nu,T_{\rm cmb})$ as function of the energy $E = h\nu$, is shown in Fig.~\ref{fig=14}. 
 Here the result is multiplied by a factor $10^{20}$ to get coincidence with the COBE data~\cite{CMB2018}.
 The upper scale is given in units of the wavenumber $k = E\hbar^{-1}c^{-1}$ ranging from 2 cm$^{-1}$ to 22 cm$^{-1}$.
 The CMB energy, $E_{\rm cmb} = k_BT_{\rm cmb}$,  is about $3.76\Times10^{-23}~{\rm J}$ or $0.235~{\rm meV}$, and the CMB wavenumber $k_{\rm cmb}$ is about $11.9~{\rm cm}^{-1}$. 
 
 One can compute the diffusion parameter 
\begin{equation}
  D_{\rm cmb} = {{R_{\rm cmb}(\nu,T_{\rm cmb})}\over{c\rho_{\rm crit}}}
\label{eq=105}
\end{equation} 
for diffusion scattering of the CMB photons in the outer space.
From here we can evaluate the diffusion length of the scattering photons
\begin{equation}
  L_{\rm cmb} =
  \sqrt{
          D_{\rm cmb} (ck)^{-1}
         }.
\label{eq=106}
\end{equation}
For $k=5.354~{\rm cm}^{-1}$ (this quantity corresponds to maximal value of the CMB radiation, see Fig.~\ref{fig=14}) the diffusion length is about $2.8\Times10^{-6}~{\rm m}$ at $\lambda = 1/k \approx 1.9\Times10^{-3}~{\rm m}$. 
It means, that the BEC is robust at such minuscule fluctuations.

Since the diffusion length is much smaller wavelengths of the CMB-photons, the pressure gradients of fluctuating virtual particles are weak, what induces weak
gravitomagnetic fluctuations. It means that the fluctuating sources ${\vec{\mathcal B}}$ and ${\vec{\mathcal E}}$ in Eqs.~(\ref{eq=98})-(\ref{eq=99}) are very weak. Nevertheless, these fluctuations bear quantum nature obeying to the Planck's law~(\ref{eq=104}).


\section{\label{sec7}Conclusion}

By beginning from the Penrose's twistor program we moved immediately from the SU(2) group to the group represented by four quaternion matrices of rank~4. This elicits the $4\pi$ symmetry as related to a non-orientable topology of the complex plane as compactified onto the Riemann sphere, as shown upon Fig.~\ref{fig=2}(b).

\begin{figure}[htb!]
  \begin{picture}(120,320)(0,0)
      \includegraphics[scale=0.65]{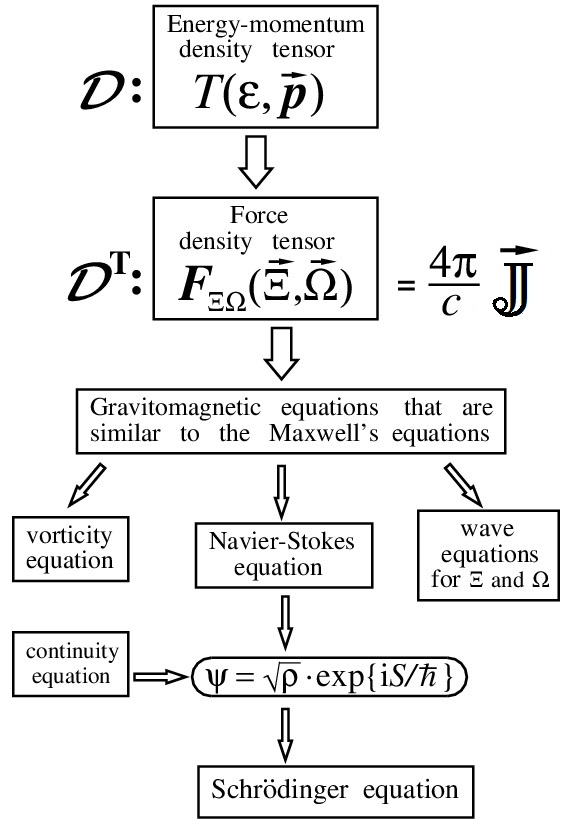}
  \end{picture}
  \caption{
Diagram showing action of the generalized differential operators ${\mathcal D}$ and ${\mathcal D}^{\bf T}$ and emergence of the vorticity equation, \Schrodinger equation, and the wave equations. 
 }
  \label{fig=15}
\end{figure} 

The diagram shown in Fig.~\ref{fig=15} summarizes the results presented in this work. 
Differential operators $\partial/\partial x$, $\partial/\partial y$, $\partial/\partial z$, and ${\bf i}\partial/c\partial t$ (${\bf i}$ is an imaginary unit  and $c$ is the speed of light),  being  coefficients of the four 4$\times$4 quaternion matrices, serve for reproduction of the generating operator~$\mathcal D$. This operator maps the energy-momentum tensor onto the force density tensor. The latter is composed of the sum of the external forces acting on the deformable medium from the outside and of the internal forces arising as a reaction on the external ones. The internal forces are represented by the pressure gradients causing the deformations of the medium and the energy dissipation conditioned by the shear viscosity of  the medium layers forced by the pressure gradients.

The deformable medium in our consideration is the superfluid quantum space. For that reason, it would seem, we should reject the viscosity. However it is not a good idea, since it leads to appearance of singularities in subsequent computations. Instead we do not reject the viscosity, but proclaim that the viscosity in the average of time is zero, but its variance is not zero.  In that case we deal with the energy exchange with the background environment.

A motion of this  superfluid quantum medium is described in the non-relativistic limit by the modified Navier-Stokes equation~\cite{Sbitnev2015b,Sbitnev2016b,Sbitnev2016c,Sbitnev2018b,Sbitnev2018c}, 
which computes the field of velocities on it. The modification deals with the internal forces mentioned above. In addition, the state field of this medium is described by the continuity equation. In pair these two equations give a full description of  a quantum object moving in this medium. The first equation describes the mobility of the quantum object near any point  in this medium. The second equation gives the probability of detecting this object in the vicinity of this point. The more accurately the velocity of the object is determined, the more uncertain its position. And vice versa.

The  ancient Greek philosopher Aristotle  assumed the existence of ether that fills all space everywhere densely. This celestial medium, named by him to be the fifth element,  in addition to four terrestrial elements (solid state, liquid, gaseous state, and plasma), is not trivial. In  traditional Indian cosmology this medium, named akasha, is "an ethereal fluid pervading the whole cosmos". The superfluid quantum medium can be perceived as a modern reincarnation of the above mentioned ethereal fluid.

In this respect, we determine the quantum potential stipulating the quantum nature of this medium. The quantum potential is equal to the internal pressure divided by the density distribution of quantum carriers in the medium -- particles with mass $m$. The carrier in this medium  moves along the Bohmian trajectory with random deviations from a medial direction  evaluated by the uncertainty principle. The diffusion coefficient is $\hbar/2m$, $\hbar$ is the reduced Planck constant.  Such a  modification leads to derivation of the \Schrodinger equation.

The second generating operator shown in  Fig.~{\ref{fig=15}} is the transposed operator ${\mathcal D}^{\rm T}$. 
It acts on the force density tensor loaded by the 4D current density printed from the right. The current density is produced by forces acting  both external and internal to the deformable medium.
The operator  ${\mathcal D}^{\rm T}$ by acting on the  force density tensor  reproduces four differential equations representing  the gravitomagnetic equations.  
They are similar to Maxwell's equations. There  is   a   small   difference, however. It lies in the difference of signs in some equations. 
It is remarkable, that Ignazio  Ciufolini and John Archibald  Wheeler knew about this difference and expressed it  in their monograph~\cite{CiufoliniWheeler1995} 
by drawing different orientations  of the gravitomagnetic and magnetic fields on the figure (compare Figs.~\ref{fig=5}(a) and~\ref{fig=5}(b). 
It should be noted, however, that the graviromagnetic equations and the Maxwell's equations are isomorphic. For that it is  sufficient to change directions either of the  gravitomagnetic field or of the magnetic induction. 

It is remarkable that the four gravitomagnetic equations are the generating equations for the set  of  equations   bearing  a  real   physical   meaning. 

The first equation simply says that the force density $\vec\Omega$ does not contain the forces resulting  from the gradients of potential energies - there are no contributions from scalar fields. 

The second equation leads to the vorticity equation. It means that the force density $\vec\Omega$ is the orbital force. 
Since  the viscosity coefficient is zero in the average of time but its variance is not zero, we get solutions of the vortex with the non-zero its core and with infinite vortex lifetime. It gives a pattern of the orbital velocity simulating  that of a spiral galaxy.

The third equation that can be extracted from the modified Navier-Stokes is equation the Hamilton-Jacobi equation loaded by the quantum potential. By combining the latter equation with the continuity equation through involvement of the wave function  written in the polar form (it is printed in the elongated oval in Fig.~\ref{fig=15}) we come to the \Schrodinger equation.

The fourth equation underlies emergence of two wave equations describing wave dynamics of the force densities $\vec\Omega$ and $\vec\Xi$.  
Both equations contain sources supporting propagation of the waves. Organization of these sources are strongly similar to organization of the 4D electromagnetic potential. Since the Universe contains enormous amounts of gravitational  material, the sources are present everywhere. In particular, by that reason the Universe can be compared with the absolutely black body radiating CMB that obeys to the Planck's law. 

\begin{acknowledgements}
 It is pleasant to note useful discussions with Marco Fedi  concerning  the Universe.
 The author thanks Mike Cavedon and Sridhar Bulusu for useful and valuable discussions, remarks, and offers.
\end{acknowledgements}

  

\bibliographystyle{spmpsci}      

\end{document}